\documentclass[12pt]{article}
\usepackage{graphicx}
\usepackage{longtable}
\makeatletter
 
  \@addtoreset{equation}{section}
\makeatother
\def\beqar {\begin{eqnarray}}
\def\eeqar {\end{eqnarray}}
\def\beq {\begin{equation}}
\def\eeq {\end{equation}}
\def\A{{\cal A}}
\def\B{{\cal B}}
\def\C{{\cal C}}

\def\F{{\cal F}}

\def\N{{\cal N}}

\def\S{{\cal S}}
\def\V{{\cal V}}

\def\al{\alpha}
\def\bt{\beta}
\def\del{\delta}

\def\ga{\gamma}

\def\ep{\epsilon}

\def\om{\omega}

\def\th{\theta}

\def\si{\sigma}

\def\zt{\zeta}

\def\d{\partial}

\def\Ad{{\dot A}}

\def\bu{{\bar u}}

\def\hf{\frac{1}{2}}

\def\<{\langle}
\def\>{\rangle}

\def\Tr{{\rm Tr}}

\def\Path{{\rm P}}

\def\cp{{\bf CP}}

\begin{document}

\begin{titlepage}
\null\vspace{-62pt} \pagestyle{empty}
\begin{center}
\vspace{1.0truein}

{\Large\bf Holonomies of gauge fields in twistor space 6: \\
\vspace{.35cm}
\hspace{-.3cm}
incorporation of massive fermions} \\

\vspace{1.0in} {\sc Yasuhiro Abe} \\
\vskip .12in {\it Cereja Technology Co., Ltd.\\
1-13-14 Mukai-Bldg. 3F, Sekiguchi \\
Bunkyo-ku, Tokyo 112-0014, Japan } \\
\vskip .07in {\tt abe@cereja.co.jp}\\
\vspace{1.3in}
\centerline{\large\bf Abstract}
\end{center}
Following the previous paper arXiv:1205.4827, we formulate an S-matrix functional for massive
fermion ultra-helicity-violating (UHV) amplitudes, {\it i.e.}, scattering
amplitudes of positive-helicity gluons and a pair of massive fermions.
The S-matrix functional realizes a massive extension
of the Cachazo-Svrcek-Witten (CSW) rules in a functional language.
Mass-dimension analysis implies that interactions among gluons
and massive fermions should be decomposed into three-point massive fermion subamplitudes.
Namely, such interactions are represented by combinations
of three-point UHV and next-to-UHV (NUHV) vertices.
This feature is qualitatively different from the
massive scalar amplitudes where the number of involving gluons
can be arbitrary.

\end{titlepage}
\pagestyle{plain} \setcounter{page}{2} 

\section{Introduction}

Recent developments on the calculation of scattering amplitudes
in gauge theories within a framework of the spinor-helicity formalism are remarkable.
Technically and practically these developments can be recapitulated
by a field theoretic prescription, either the CSW rules \cite{Cachazo:2004kj} or the
BCFW recursion relation \cite{Britto:2004ap,Britto:2005fq}.

Incorporation of massive fermions along the lines of these developments,
a main subject of the present paper, has been studied in various methods,
{\it e.g.}, by extension of BCFW-like recursion relations
\cite{Badger:2005jv,Rodrigo:2005eu,Ferrario:2006np,Hall:2007mz}
(see also recent progress \cite{Chen:2011sba,Britto:2012qi}),
by use of supersymmetric Ward identities \cite{Schwinn:2006ca,Huang:2012gs},
and by developing massive versions of the CSW rules \cite{Ettle:2008ey,Schwinn:2008fm}.
(The massive CSW rules are initially developed in \cite{Boels:2007pj,Boels:2008ef}
for the incorporation of massive scalars into the original CSW rules.)
There also exists a more general approach
which seems to connect all these methods to one another;
this is referred to as the ``on-shell constructibility'' method
\cite{Cohen:2010mi,Boels:2011zz}.
In this general approach massive deformation of massless amplitudes
is carried out by use of the massive spinor-helicity formalism \cite{Dittmaier:1998nn}.
The on-shell constructibility method is applicable to massive fermions
as well as the other types of massive particles, {\it i.e.}, scalars and vector bosons
\cite{Cohen:2010mi,Boels:2011zz}.

In a recent paper \cite{Abe:2012en} we construct an S-matrix functional for
massive scalar amplitudes, {\it i.e.}, amplitudes of an arbitrary number of gluons
and a pair of massive scalar particles, in
the recently proposed holonomy formalism \cite{Abe:2009kn,Abe:2009kq}.
In \cite{Abe:2012en} we show that the on-shell constructibility
of massive scalar amplitudes can naturally be implemented by means of
off-shell continuation of Nair's superamplitude method \cite{Nair:1988bq}
once a massive holonomy operator is defined.
We also show that a careful study of the massive holonomy operator
leads to a novel color structure of the massive scalar amplitudes.
A main purpose of the present paper is to extend these results
to the case of massive fermions in the same framework.

The organization of this paper is as follows.
In section 2 we review essential results of the
previous paper \cite{Abe:2012en} on the massive scalar amplitudes
in the holonomy formalism.
We first write down the definitions of the massless and massive holonomy operators.
We then present an S-matrix functional for the massive scalar tree amplitudes,
focusing on the amplitudes of positive-helicity gluons and a pair of
massive scalars, the so-called ultra-helicity-violating (UHV)
massive scalar amplitudes.

In section 3 we apply these results to the massive fermion amplitudes.
We begin with clarifying our notation of fermions
in the spinor-helicity formalism. We then apply off-shell continuation
of Nair's superamplitude method such that it lead to tree-level
massive fermion UHV amplitudes (scattering amplitudes
of an arbitrary number of positive-helicity gluons
and a pair of massive fermions) in a form that is consistent
with the literature. We also obtain an S-matrix functional for
the massive fermion UHV amplitudes.

Mass-dimension analysis on the massive fermion UHV amplitudes
implies that the number of gluons involving the amplitudes
should be one.
This condition means that interactions among gluons
and massive fermions should be decomposed into the
three-point UHV vertices
(with a positive-helicity gluon) and
the three-point next-to-UHV (NUHV) vertices
(with a negative-helicity gluon).
In section 4, for clarification of our arguments, we
first construct the massive fermion NUHV amplitudes
for arbitrary number of gluons and see how the above
constraint is imposed.
Lastly, we present a brief conclusion.

\section{S-matrix for massive scalar UHV amplitudes}

In this section we review the formulation of
an S-matrix functional for the massive scalar UHV amplitudes, {\it i.e.},
scattering amplitudes of an arbitrary number of positive-helicity gluons and
a pair of massive scalar particles, at tree level in the framework of
the holonomy formalism, recapitulating the results of the
recent study \cite{Abe:2012en}.

\noindent
\underline{The massless holonomy operator}

The original ``massless'' holonomy operator for gluons is defined as \cite{Abe:2009kn}
\beq
    \Theta_{R, \ga}^{(A)} (u) = \Tr_{R, \ga} \, \Path \exp \left[
    \sum_{m \ge 2} \oint_{\ga} \underbrace{A \wedge A \wedge \cdots \wedge A}_{m}
    \right]
    \label{2-1}
\eeq
where $A$ is called the comprehensive gluon field. This is expressed as a bialgebraic
operator
\beqar
    A & = & g \sum_{1 \le i < j \le n} A_{ij} \, \om_{ij} \, ,
    \label{2-2} \\
    A_{ij} & = & a_{i}^{(+)} \otimes a_{j}^{(0)} + a_{i}^{(-)} \otimes a_{j}^{(0)} \, ,
    \label{2-3} \\
    \om_{ij} & = & d \log( u_i u_j ) \, = \, \frac{d( u_i u_j )}{( u_i u_j )}
    \label{2-4}
\eeqar
where $g$ is a dimensionless coupling constant
and $u_i$  ($i = 1,2, \cdots , n$) denotes
the two-component spinor momentum for the $i$-th gluon.
$(u_i u_j )$ represents a scalar product of the spinor momenta:
\beq
    (u_i u_j) \, = \, \ep_{AB} u_{i}^{A}u_{j}^{B} \, \equiv  \, u_i \cdot u_j
    \label{2-5}
\eeq
where the indices $A,B$ take a value of $(1,2)$
and $\ep_{AB}$ denotes the rank-2 Levi-Civita tensor.
The physical (creation) operators of positive and negative helicity gluons
are given by $a_{i}^{(+)}$ and $a_{i}^{(-)}$ in (\ref{2-3}), respectively.
These form the ladder part of the $SL ( 2, {\bf C} )$ algebra, satisfying
\beq
    [ a_{i}^{(+)}, a_{j}^{(-)}] = 2 a_{i}^{(0)} \, \del_{ij}  \, , ~~~
    [ a_{i}^{(0)}, a_{j}^{(+)}] = a_{i}^{(+)} \, \del_{ij} \, , ~~~
    [ a_{i}^{(0)}, a_{j}^{(-)}] = - a_{i}^{(-)} \, \del_{ij}
    \label{2-6}
\eeq
where Kronecker's deltas show that
the non-zero commutators are obtained only when $i = j$.
The remaining commutators, those expressed otherwise, all vanish.

The color degree of freedom
can be attached to the physical operators $a_{i}^{(\pm)}$  as
\beq
    a_{i}^{(\pm)} = t^{c_i} \, a_{i}^{(\pm)c_i}
    \label{2-7}
\eeq
where $t^{c_i}$'s are the generators of the $U(N)$ gauge group in the $R$-representation.
The symbol $\Path$ in (\ref{2-1}) denotes an ordering of the numbering indices.
The meaning of the
action of $\Path$ on the exponent of (\ref{2-1}) can be written explicitly as
\beqar
    &&
    \Path \sum_{m \ge 2}  \oint_{\ga} \underbrace{A \wedge \cdots \wedge A}_{m}
    \nonumber \\
    &=& \sum_{m \ge 2} \oint_{\ga}  A_{1 2} A_{2 3} \cdots A_{m 1}
    \, \om_{12} \wedge \om_{23} \wedge \cdots \wedge \om_{m 1}
    \nonumber \\
    &=& \sum_{m \ge 2}  \frac{1}{2^{m+1}} \sum_{(h_1, h_2, \cdots , h_m)}
    (-1)^{h_1 + h_2 + \cdots + h_m} \,
    a_{1}^{(h_1)} \otimes a_{2}^{(h_2)} \otimes \cdots \otimes a_{m}^{(h_m)}
    \, \oint_{\ga} \om_{12} \wedge \cdots \wedge \om_{m1}
    \nonumber \\
    \label{2-8}
\eeqar
where $h_{i} = \pm = \pm 1$ ($i=1,2,\cdots, m$) denotes
the helicity of the $i$-th gluon.

In (\ref{2-1}), $\ga$ represents a closed path on the physical configuration space
\beq
    \C^{(A)} = {\bf C}^n / \S_n
    \label{2-9}
\eeq
on which the set of gluon operators are defined.
$\S_n$ denotes the rank-$n$ symmetric group.
The fundamental homotopy group of $\C^{(A)}$ is
given by the braid group, $\Pi_1 ( \C^{(A)} ) = \B_n$.
The trace $\Tr_{\ga}$ in (\ref{2-1}) is taken over the
the generators of the braid group (or the elements of Iwahori-Hecke algebra,
see \cite{Abe:2009kq} for details on this point) and is
called the braid trace.
It can be realized by a sum over the permutations of the numbering indices,
{\it i.e.},
\beq
    \Tr_{\ga} \, = \, \sum_{\si \in \S_{n-1} }
    \label{2-10}
\eeq
where each of the permutations is denoted by
$\si=\left(%
\begin{array}{c}
  2 ~ 3 \, \cdots \, n \\
  \si_2 \si_3 \cdots \si_n \\
\end{array}%
\right)$.
Explicitly, the braid trace $\Tr_\ga$ over the exponent (\ref{2-8}) can be expressed as
\beq
    \Tr_{\ga} \Path \sum_{m \ge 2}^{\infty} \oint_{\ga}
    \underbrace{A \wedge \cdots \wedge A}_{m}
    = \sum_{m \ge 2}
    \sum_{\si \in \S_{m-1}} \oint_{\ga}  A_{1 \si_2} A_{\si_2 \si_3} \cdots A_{\si_m 1}
    \, \om_{1 \si_2} \wedge \om_{\si_2 \si_3} \wedge \cdots \wedge \om_{\si_m 1} \, .
    \label{2-11}
\eeq

A quintessential point in the holonomy formalism is the mathematical fact
that a linear representation of a braid group is equivalent to
a monodromy representation of the Knizhnik-Zamolodchikov (KZ) equation.
Such a monodromy representation can be given by a holonomy of
the so-called KZ connection. The comprehensive gauge one-form (\ref{2-2})
is a variant of the KZ connection in a sense that it satisfies
the infinitesimal braid relations to ensure the integrability of the
comprehensive gauge one-form. For details on these fundamental issues,
see \cite{Abe:2009kn,Abe:2009kq}; see also
\cite{Cirio:2011ja,Cirio:2012be} for recent studies on
KZ connections and two-dimensional holonomies.

\noindent
\underline{The massive holonomy operator}

Following these basic ideas, we now construct a holonomy operator for
a comprehensive gauge one-form which incorporates the helicity-zero
(or spin-zero) scalar operators $a_{i}^{(0)}$.
Such a one-form can be defined as
\beqar
    B & = & \sum_{1 \le i < j \le n} B_{ij} \, \om_{ij} \, ,
    \label{2-12} \\
    B_{ij} & = & g \left(
    a_{i}^{(+)} \otimes a_{j}^{(0)} +  a_{i}^{(-)} \otimes a_{j}^{(0)}
    \right)
    \, + \,
    a_{i}^{(0)} \otimes a_{j}^{(0)}
    \label{2-13}
\eeqar
where $g$ denotes the dimensionless gauge coupling constant as before.
The holonomy operator for $B$ is then defined and calculated as \cite{Abe:2012en}
\beqar
    \Theta_{R, \ga}^{(B)} (u) &=& \Tr_{R, \ga} \, \Path \exp \left[
    \sum_{r \ge 2} \oint_{\ga} \underbrace{B \wedge B \wedge \cdots \wedge B}_{r}
    \right]
    \nonumber \\
    &=& \exp \left[
    \sum_{r \ge 3}
    \sum_{ ( h_2 , h_3 , \cdots , h_{r-1} ) }
    g^{r-2} ( -1 )^{ h_2 h_3 \cdots h_{r-1} }
    \,
    \Tr \left(
    t^{c_2} t^{c_3} \cdots t^{c_{r-1}}
    \right)
    \right.
    \nonumber \\
    &&
    \left.
    ~~~~~~~ \times \,
    \frac{
    a_{1}^{(0)} \otimes a_{2}^{(h_2)c_2} \otimes \cdots \otimes a_{r-1}^{(h_{r-1})c_{r-1}}
    \otimes a_{r}^{(0)}
    }{
    (12)(23) \cdots (r-1 \, r)( r 1)
    }
    \right]
    \label{2-14}
\eeqar
where we use the abbreviated notation $( i \, j ) \equiv (u_i u_j ) $.
In the above expression the massive scalar operators are specified by
a pair of the numbering indices $(1, r)$.
Accordingly $h_{i} = \pm = \pm 1$ ($i=2,3, \cdots, r-1$) denotes
the helicity of the $i$-th gluon.

For an $n$-particle system, the physical configuration space is given by
\beq
    \C^{(B)} \, = \, \frac{ {\bf C}^{n-2} }{ \S_{n-2} }  \otimes {\bf C}^2
    \, = \,  {\bf C}^{n} / \S_{n-2}
    \label{2-15}
\eeq
as opposed to the purely gluonic case in (\ref{2-9}).
Consequently the braid trace $\Tr_\ga$ is taken over the
numbering elements $\{ \si_2 , \si_3 , \cdots , \si_{r-1} , \tau_r \}
= \{ 2,3,\cdots , r \}$, satisfying the $\Path$ ordering
\beq
    \si_2 < \si_3 < \cdots < \si_{r-1} \, .
    \label{2-16}
\eeq
The braid trace $\Tr_\ga$ in (\ref{2-14}) can then be represented by a ``homogenous'' sum
\beq
    \sum_{ \{ \si , \tau \} }
    \, = \,
    \sum_{\tau_r = 2}^{r} \, \sum_{\si \in \S_{r-2} }
    \label{2-17}
\eeq
where $r = 3, 4, \cdots , n$.

\noindent
\underline{Comments on the color structure of the massive holonomy operator}

As mentioned in the introduction, the color structure of the
massive holonomy operator is qualitatively
different from that of the massless holonomy operator
in terms of the final realization of the braid trace.
In the original massless case, the sum over the permutations of the
numbering indices (\ref{2-10}) is explicit while in
the massive case, as shown in (\ref{2-14}), there
appears no explicit sums over permutations.
As discussed in \cite{Abe:2012en}, this is due to the fact
that the calculation of (\ref{2-14}) is carried out by use of the relation
\beq
    \sum_{ \{ \si , \tau \} } \oint_\ga
    \om_{1 \si_2} \wedge \om_{\si_2  \si_3} \wedge \cdots \wedge
    \om_{\si_{r-1}  \tau_{r} } \wedge \om_{ \tau_{r} 1}
    \, = \,
    \int_{\ga_{1 r}} \om_{12} \wedge \cdots \wedge \om_{r-1 \, r}
    \int_{\ga_{r 1}} \om_{r1}
    \label{2-18}
\eeq
where $\ga_{1 r}$ and $\ga_{r 1}$
denote open paths on $\C^{(B)}$  which
compose the closed path, $\ga = \ga_{1r} \ga_{r1}$.

The relation (\ref{2-18}) is a mathematical equation regarding
products of iterated integrals over the logarithmic one-forms $\om_{ij}$'s.
Use of this relation implies that the homogeneous sum (\ref{2-17}) or the
braid trace for $\Theta_{R, \ga}^{(B)} (u)$ can be absorbed into
the expression of (\ref{2-14}).

\noindent
\underline{Supersymmetrization}

In order to relate the massive holonomy operator to
massive scalar amplitudes, we need to define the massive holonomy operator
in supertwistor space, as in the case of gluon amplitudes.
In the massless case the supersymmetrization is carried out by
replacing the physical operators $a_{i}^{(\pm )}$ in (\ref{2-3})
with
\beq
    a_{i}^{( \pm )} (x, \th) \, = \,
    \left. \int d \mu (p_i) ~ a_{i}^{( \pm )} (\xi_i) ~  e^{ i x_\mu p_{i}^{\mu} }
    \right|_{\xi_{i}^{\al} = \th_{A}^{\al} u_{i}^{A} }
    \label{2-19}
\eeq
where $d \mu (p_i)$ is the ordinary Lorentz-invariant measure for the null momentum $p_i^\mu$.
This is also called the Nair measure when it is written in terms of the spinor momenta
\cite{Nair:1988bq}.
The operators $a_{i}^{(\hat{h}_{i})} (x, \th)$ are physical operators
that are defined in a four-dimensional $\N = 4$ chiral
superspace $(x, \th)$ where $x_{\Ad A} = (\si^\mu )_{\Ad A} x_\mu$
denote coordinates of four-dimensional spacetime and $\th_{A}^{\al}$ $(A = 1,2; \al = 1,2,3,4)$
denote their chiral superpartners with $\N = 4$ extended supersymmetry.
(Here $\si^\mu$ is given by $\si^\mu = ({\bf 1}, \si^i )$, with $\si^i$ ($i = 1,2,3$)
being the Pauli matrices.)
These coordinates emerge from homogeneous coordinates
of the supertwistor space $\cp^{3|4}$, represented by $( u^A , v_\Ad , \xi^\al )$,
that satisfy the so-called supertwistor conditions
\beq
    v_\Ad \, = \, x_{\Ad A} u^A \, , ~~~
    \xi^\al \, = \, \th_{A}^{\al} u^A \, .
    \label{2-20}
\eeq

Upon the supersymmetrization, there arise superpartners of the gluon
creation operators. The full supermultiplets can be expressed as
\beqar
    \nonumber
    a_{i}^{(+)} (\xi_i) &=& a_{i}^{(+)} \, ,
    \\
    \nonumber
    a_{i}^{\left( + \frac{1}{2} \right)} (\xi_i) &=& \xi_{i}^{\al}
    \, a_{i \, \al}^{ \left( + \frac{1}{2} \right)} \, ,
    \\
    a_{i}^{(0)} (\xi_i) &=& \hf \xi_{i}^{\al} \xi_{i}^{\bt} \, a_{i \, \al \bt}^{(0)}
    \, ,
    \label{2-21}
    \\
    \nonumber
    a_{i}^{\left( - \frac{1}{2} \right)} (\xi_i) &=&
    \frac{1}{3!} \xi_{i}^{\al}\xi_{i}^{\bt}\xi_{i}^{\ga}
    \ep_{\al \bt \ga \del} \, a_{i}^{ \left( - \frac{1}{2} \right) \, \del}
    \, ,
    \\
    \nonumber
    a_{i}^{(-)} (\xi_i) &=& \xi_{i}^{1} \xi_{i}^{2} \xi_{i}^{3} \xi_{i}^{4} \, a_{i}^{(-)} \, .
\eeqar
These operators are characterized by the number of $\xi_i^\al = \th_{A}^{\al} u^A_i$
or the degrees of homogeneities in $u^A_i$'s.
The number is in one-to-one correspondence with the helicity
$\hat{h}_{i} = (0, \pm \hf , \pm 1 )$ of the superpartner of interest.
This can be easily seen from the definition of the helicity operator
\beq
    \hat{h}_{i} = 1 - \hf  u_{i}^{A} \frac{\d}{\d u_{i}^{A}} \, .
    \label{2-22}
\eeq
Notice that we put the hat of $\hat{h}_i$ simply to distinguish it
from the non-supersymmetric version $h_i = \pm $.

Use of the supermultiplets (\ref{2-21}) enables us to
define gluon operators without introducing the conventional polarization/helicity vectors.
This method is known as Nair's superamplitude method \cite{Nair:1988bq}.
In \cite{Abe:2012en} we show that this prescription is
also applicable to the system of gluons and a pair of massive scalars.
In the following we review such a massive extension.

\noindent
\underline{Off-shell continuation of Nair's superamplitude method}

In carrying out the supersymmetrization of the massive holonomy operator
(\ref{2-14}), we need to consider off-shell version of the
scalar operator $a_{i}^{(0)} ( \xi_i )$ in (\ref{2-21}).
In the massive spinor-helicity formalism \cite{Dittmaier:1998nn},
off-shell continuation of the null spinor-momentum $u^A$ is
defined as
\beq
    u^A ~ \longrightarrow ~
    \widehat{u}^A \, = \,  u^A + \frac{m}{( u \eta ) } \eta^A
    \label{2-23}
\eeq
where $\eta^A$ is a reference null spinor and $m$ denotes the
mass of the massive spinor momentum $\widehat{u}^A$.
The complex conjugate of (\ref{2-23}) is also necessary to
define the full massive four-momentum $\widehat{p}^{A \Ad} =
\widehat{u}^A \widehat{\bu}^\Ad$.
We then introduce another set of supertwistor variables
$( w^A , \pi_\Ad , \zt^\al )$ such that
the supertwistor conditions
\beq
    \pi_\Ad \, = \, x_{\Ad A} w^A \, = \, x_{\Ad A} \frac{m}{( u \eta ) } \eta^A
    \, , ~~~
    \zt^\al \, = \, \th_{A}^{\al} w^A
    \, = \, \th_{A}^{\al} \frac{m}{( u \eta ) } \eta^A
    \label{2-24}
\eeq
are satisfied ($\al = 1,2,3,4)$.

Using the new Grassmann variable $\zt^\al_i = \th_{A}^{\al}
\frac{m}{( u_i \eta_i ) } \eta_i^A$, the off-shell continuation of the scalar operator
$a_{i}^{(0)} ( \xi_i )$ can be defined as
\beq
    a_{i}^{(0)} ( \xi_i ) ~ \longrightarrow ~
    a_{i}^{(0)} ( \xi_i , \zt_i ) ~ = ~
    \xi_{i}^{1} \xi_{i}^{2} \xi_{i}^{3} \zt_{i}^{4} \, a_{i}^{(0)}
    \label{2-25}
\eeq
where we shall specify the numbering index to $i = 1, n$ for massive scalars.
Notice that the degrees of homogeneities in $u_i$'s remains the same in
the off-shell continuation (\ref{2-25}). Thus we can naturally interpret
$a_{i}^{(0)} ( \xi_i , \zt_i )$ as massive scalar operators.
The chiral superspace representation of the massive operators
can be expressed as
\beqar
    a_{i}^{( 0 )} (x, \th)  & = &
    \left. \int d \mu ( \widehat{p}_i ) ~ a_{i}^{(0)} ( \xi_i , \zt_i )
    ~  e^{ i x_\mu \widehat{p}_i^\mu }
    \right|_{ \xi_{i}^{\al} = \th_{A}^{\al} u_i^A ,
    \, \zt_{i}^{\al} = \th_{A}^{\al} w_i^A  }
    \, ,
    \label{2-26} \\
    \widehat{p}_i^\mu & = & p_{i}^{\mu} + \frac{m^2}{2 (p_i \cdot \eta_i )} \eta_i^\mu
    \, ,
    \label{2-27} \\
    w_i^A & = &   \frac{m}{( u_i \eta_i ) } \eta_i^A
    \, .
    \label{2-28}
\eeqar

A supersymmetric version of the massive holonomy operator can be constructed
by use of the the expressions (\ref{2-19}) and (\ref{2-26})
for physical operators of  gluons and massive scalars, respectively.
{\it
In other words, we can obtain
the supersymmetric massive holonomy operator $\Theta_{R , \ga}^{(B)} (u; x, \th)$
out of $\Theta_{R , \ga}^{(B)} (u)$ in (\ref{2-14}) by replacing
$\{ a_{i}^{(\pm )} , a_{j}^{( 0 )} \}$ with
$\{ a_{i}^{(\pm )}  ( x, \th ) , a_{j}^{(0)}  ( x, \th ) \}$
where $i=2,3,\cdots , r-1$ and $j = 1, r$.
}

\noindent
\underline{Choice of reference spinors and functional derivation of the UHV vertex}

We now specify the reference null-vectors for the massive scalars as
\beq
    \eta_1^\mu = p_n^\mu \, , ~~~~ \eta_n^\mu = p_1^\mu \, .
    \label{2-29}
\eeq
Namely, we choose a specific parametrization of
$\widehat{p}_1$ and $\widehat{p}_n$ by
\beq
    \widehat{p}_1^\mu \, = \, p_1^\mu + \frac{m^2}{2 (p_1 \cdot p_n )} p_n^\mu
    \, , ~~~
    \widehat{p}_n^\mu \, = \, p_n^\mu + \frac{m^2}{2 (p_n \cdot p_1 )} p_1^\mu
    \, .
    \label{2-30}
\eeq
Notice that this parametrization is qualitatively different from
off-shell prescription for virtual gluons where we set all reference null-vectors
identical.

We now introduce a generating functional
\beq
    \F_{\rm UHV}^{\rm (vertex)} \left[ a^{( \pm )c} , a^{(0)} \right]
    \, = \,
    \exp \left[
    i \int d^4 x d^8 \th \, \Theta_{R , \ga}^{(B)} (u; x, \th)
    \right] \, .
    \label{2-31}
\eeq
In terms of this generating functional the massive scalar UHV vertex can be
obtained as
\beqar
    &&
    \V_{\rm UHV}^{( \bar{\phi}_{1} g_{2}^{+} \cdots g_{n-1}^{+}  \phi_{n} )} (x)
    \nonumber \\
    &=&
    \left.
    \frac{\del}{\del a_{1}^{(0)}} \otimes
    \frac{\del}{\del a_{2}^{(+)}} \otimes
    \frac{\del}{\del a_{3}^{(+)}} \otimes \cdots \otimes
    \frac{\del}{\del a_{n-1}^{(+)}} \otimes
    \frac{\del}{\del a_{n}^{(0)}}
    \, \F_{\rm UHV}^{\rm (vertex)} \left[ a^{( \pm )c} , a^{(0)} \right]
    \right|_{a^{( \pm )c} = a^{(0)} = 0}
    \label{2-32}
\eeqar
where $a^{( \pm )c}$ and $a^{(0)}$ denote sets of source functions
of gluons and massive scalars, respectively.
$\V_{\rm UHV}^{( \bar{\phi}_{1} g_{2}^{+} \cdots g_{n-1}^{+}  \phi_{n} )} (x)$
denotes the massive scalar UHV vertex in the $x$-space representation.
Explicitly these are expressed as
\beqar
    \V_{\rm UHV}^{( \bar{\phi}_{1} g_{2}^{+} \cdots g_{n-1}^{+}  \phi_{n} )} (x)
    & \equiv &
    \V_{\rm UHV}^{( \bar{\phi}_1 \phi_n )} (x)
    ~ = ~
    \int d \mu (\widehat{p}_1 )
    \prod_{i=2}^{n-1}  d \mu ( p_i )
    d \mu (\widehat{p}_n )
    \, \V_{\rm UHV}^{( \bar{\phi}_1 \phi_n )} ( u , \bu ) \, ,
    \label{2-33} \\
    \V_{\rm UHV}^{( \bar{\phi}_1 \phi_n  )} ( u , \bu )
    & = &
    - i g^{n-2}
    \, (2 \pi)^4 \del^{(4)} \left( \widehat{p}_1 +
    \sum_{i=2}^{n-1} p_i + \widehat{p}_n \right) \,
    \widehat{V}_{\rm UHV}^{( \bar{\phi}_1  \phi_n )} (u)
    \, ,
    \label{2-34} \\
    \widehat{V}_{\rm UHV}^{( \bar{\phi}_1 \phi_n )} (u)
    & = &
    \Tr ( t^{c_{2}} t^{c_{3}} \cdots t^{c_{n-1}}) \,
    \frac{ m^2 \, (n 1)^2}{ (12)(23) \cdots (n-1 \, n) (n 1) } \, .
    \label{2-35}
\eeqar
The integral measures in (\ref{2-33}) are given by the ordinary
Lorentz invariant measures
\beqar
    &&
    d \mu ( \widehat{p}_1 ) \, = \, \frac{d^3 \widehat{p}_1 }{(2 \pi)^3}
    \frac{1}{2 \sqrt{|\vec{\widehat{p}}_{1}|^{2} + m^2 }}
    \, , ~~~
    d \mu ( p_i ) \, = \, \frac{d^3 p_i }{(2 \pi)^3}
    \frac{1}{2 \sqrt{|\vec{p}_{i}|^{2} }}
    \, ,
    \nonumber \\
    &&
    d \mu ( \widehat{p}_n ) \, = \, \frac{d^3 \widehat{p}_n }{(2 \pi)^3}
    \frac{1}{2 \sqrt{|\vec{\widehat{p}}_{n}|^{2} + m^2 }}
    \, .
    \label{2-36}
\eeqar
The momenta of the massive scalars are denoted by $\widehat{p}_1$ and
$\widehat{p}_n$, satisfying
\beq
    \widehat{p}_1^\mu \widehat{p}_{1 \mu}
    \, = \,     \widehat{p}_n^\mu \widehat{p}_{n \mu} \, = \, m^2 \, .
    \label{2-37}
\eeq

In deriving the explicit form (\ref{2-35}), we use the specified
reference spinors (\ref{2-29}) and the Grassmann integral
\beqar
    &&
    \int d^8 \th \, \xi^1_1 \xi^2_1 \xi^3_1 \zt^4_1
    \, \xi^1_n \xi^2_n \xi^3_n \zt^4_n \,
    \nonumber \\
    &=&
    (1 n)^3 \int d^2 \th \, \frac{m}{(u_1 \eta_1)} \eta_1^A \th_A \,
    \frac{m}{(u_n \eta_n )} \eta_n^B \th_B
    \nonumber \\
    &=&
    (1 n)^3 \frac{m^2}{( u_1 u_n ) ( u_n u_1 )} \underbrace{
    \int d^2 \th \, u_n^A \th_A \, u_1^B \th_B
    }_{= \, ( u_n u_1)}
    ~ = ~
    m^2 \, (n 1)^2 \, .
    \label{2-38}
\eeqar
This Grassmann integral guarantees that only the UHV-type vertices
survive upon the evaluation of functional derivatives in (\ref{2-32}).
This structure is analogous to the CSW rules, or the MHV rules,
in the calculation of gluon amplitudes \cite{Cachazo:2004kj}.

\noindent
\underline{The UHV rules and the massive scalar UHV tree amplitudes}

The UHV rules can be understood as a massive extension
of the MHV rules to the massive scalar UHV amplitudes. Namely,
the UHV rules state that the massive scalar UHV amplitudes
can be obtained by connecting the UHV vertices
with massive scalar propagators.
This is analogous to the MHV rules in structure
but a main difference exists, that is,
for the massive scalar UHV amplitudes we
have freedom to connect an arbitrary number of the
UHV vertices since such procedure, contrary to
the MHV rules, does not
alter the overall helicity configuration of the UHV amplitudes.
In compensation, we no longer take a sum over permutations of gluons
for the massive UHV amplitudes and non-UHV amplitudes in general.
This peculiar nature of the massive amplitudes is due to
the definition of massive holonomy operator (\ref{2-14}).

\begin{figure} [htbp]
\begin{center}
\includegraphics[width=150mm]{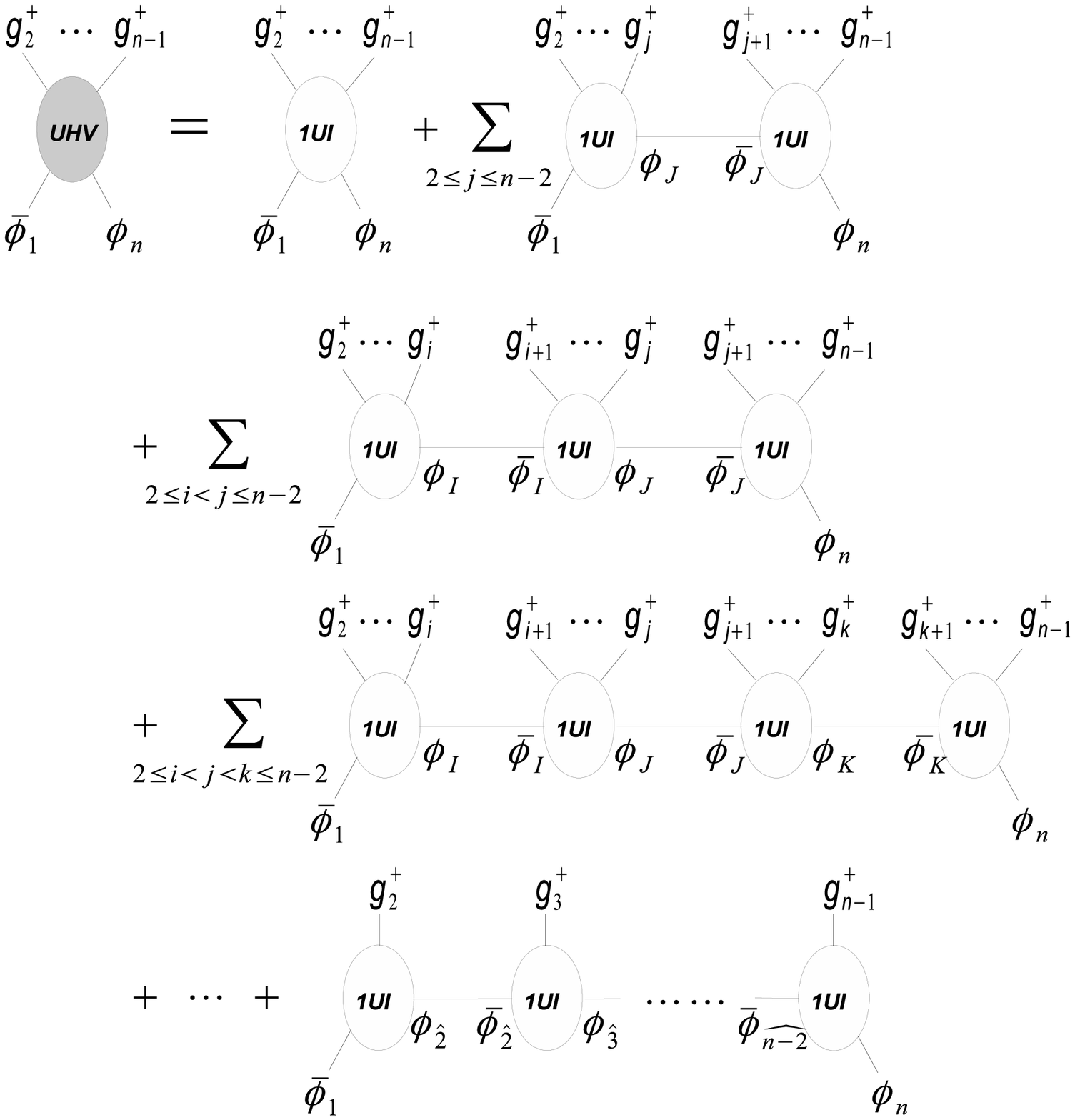}
\caption{The UHV rules for the massive scalar UHV amplitudes --- ``1UI'' stands
for a 1 UHV irreducible diagram which is equivalent to the massive scalar UHV
vertex. The vertices are connected with one another by massive scalar propagators.}
\label{fighol0601}
\end{center}
\end{figure}

Probably the easiest way to understand the UHV rules
is to visualize the full expression for the massive scalar UHV amplitudes,
see Figure \ref{fighol0601}.
As mentioned above, we have a freedom to arbitrarily add UHV vertices to
the UHV amplitudes as long as the total number of the scattering
particles is preserved.
In this sense, the UHV vertices can be interpreted as
1 UHV irreducible (1UI) subamplitudes. This is why
we denote each of the UHV vertices
by ``1UI'' in  Figure \ref{fighol0601}.

Using the UHV rules, we can then obtain an explicit form of the massive
scalar UHV amplitudes as follows.
\beqar
    \A_{\rm UHV}^{( \bar{\phi}_{1} g_{2}^{+} \cdots g_{n-1}^{+}  \phi_{n} )} (x)
    & \equiv &
    \A_{\rm UHV}^{( \bar{\phi}_1 \phi_n )} (x)
    ~ = ~
    \int d \mu (\widehat{p}_1 )
    \prod_{i=2}^{n-1}  d \mu ( p_i )
    d \mu (\widehat{p}_n )
    \, \A_{\rm UHV}^{( \bar{\phi}_1 \phi_n )} ( u , \bu ) \, ,
    \label{2-39}\\
    \A_{\rm UHV}^{( \bar{\phi}_1 \phi_n )} ( u , \bu )
    &=&
    - i g^{n-2}
    \, (2 \pi)^4 \del^{(4)} \left( \widehat{p}_1 +
    \sum_{i=2}^{n-1} p_i + \widehat{p}_n \right) \,
    \widehat{A}_{\rm UHV}^{(  \phi_n \bar{\phi}_1 )} (u)
    \, ,
    \label{2-40} \\
    \widehat{A}_{\rm UHV}^{( \bar{\phi}_1  \phi_n )} (u)
    & = &
    \Tr ( t^{c_{2}} t^{c_{3}} \cdots t^{c_{n-1}}) \,
    \widehat{C}_{\rm UHV}^{(  \bar{\phi}_1 \phi_n  )} (u)
    \, .
    \label{2-41}
\eeqar
The color-stripped holomorphic
massive scalar UHV amplitudes
$\widehat{C}_{\rm UHV}^{( \bar{\phi}_1  \phi_n )} (u)$
is given by
\beq
    \widehat{C}_{\rm UHV}^{ ( \bar{\phi}_{1}  \phi_{n} ) } ( u )
    \, = \,
    \widehat{C}_{\rm UHV}^{ ( \bar{\phi}_{1} g_{2}^{+} g_{3}^{+}
    \cdots g_{n-1}^{+} \phi_{n} ) } ( u )
    \, = \,
    \frac{ m^2 (n1) }{(12)(23)\cdots (n1)}
    \widehat{(n1)}
    \label{2-42}
\eeq
where
\beqar
    \widehat{(n1)}
    &=&
    (n1) \, + \, \sum_{j =2}^{n-2}
    \frac{(J1)}{(jJ)}
    \frac{ m^2 (j \, j+1)}{ \widehat{P}_J^2 - m^2}
    \frac{(nJ)}{(J\, j+1)}
    \nonumber \\
    &&
    + \,
    \sum_{2 \le i < j \le n-1 }
    \frac{(I1)}{(iI)}
    \frac{ m^2 (i \, i+1)}{ \widehat{P}_I^2 - m^2}
    \frac{(JI)}{(I \, i+1)(j J)}
    \frac{ m^2 (j \, j+1)}{ \widehat{P}_J^2 - m^2}
    \frac{(nJ)}{(J \, j+1)}
    \nonumber \\
    &&
    + \, \sum_{2 \le i < j < k \le n-1 }  \, \Biggl[ \,
    \frac{(I1)}{(iI)}
    \frac{ m^2 (i \, i+1)}{ \widehat{P}_I^2 - m^2}
    \frac{(JI)}{(I \, i+1)(j J)}
    \frac{ m^2 (j \, j+1)}{ \widehat{P}_J^2 - m^2}
    \nonumber \\
    &&
    ~~~~~~~~~~~~~~~~~~~~~~~~~~~
    \times \,
    \frac{(KJ)}{(J \, j+1)(k K)}
    \frac{ m^2 (k \, k+1)}{ \widehat{P}_K^2 - m^2}
    \frac{(nK)}{(K \, k+1)}
    \,   \Biggr]
    \nonumber \\
    &&
    + \, \cdots \,
    \nonumber \\
    &=&
    - \left( 1 \left|
    \prod_{j = 2}^{n-2} \left[
    1 - \frac{  m^2 |J ) ( j \, j+1 ) (J | }
    {(\widehat{P}_J^2 - m^2) (j J) (J \, j+1 ) }
    \right]
    \right| n \right) \, .
    \label{2-43}
\eeqar
In the above expression, $(1 | n ) = (1 n) = \ep_{AB} u_1^A u_n^B$
and the uppercase letters label a set of momenta running along
each of the propagators. For example, $\widehat{P_J}^\mu$ is defined as
\beq
    \widehat{P_J}^\mu  \, = \, \widehat{p}_1^\mu + p_2^\mu + p_3^\mu
    + \cdots + p_j^\mu
    \, = \,
    p_J^\mu + \frac{m^2}{2 ( p_J \cdot \eta_J ) } \eta_J^\mu
    \label{2-44}
\eeq
where $p_J^\mu$ is the on-shell partner of the off-shell momentum $\widehat{P_J}$,
$\eta_J^\mu$ is the corresponding reference null-vector, and $m$ denotes the mass
of the complex scalar particles $( \bar{\phi}_1 , \phi_n )$.
The spinor momenta $(u_J^A , \bu_J^\Ad )$ correspond to
the null momentum $p_J^\mu$ and are defined as usual:
\beq
    p_{J}^{A \Ad} \, = \, u_J^A \bu_J^\Ad
    \, .
    \label{2-45}
\eeq

\noindent
\underline{An S-matrix functional for the massive scalar UHV tree amplitudes}

It is now straightforward to introduce an S-matrix functional for
the massive scalar UHV tree amplitudes (\ref{2-39})
by use of
the supersymmetric massive holonomy operator $\Theta_{R , \ga}^{(B)} (u; x, \th)$:
\beqar
    \F_{\rm UHV}  \left[ a^{(h)c}, a^{(0)} \right]
    & = &
    \widehat{W}^{(0)} (x) \,
    \exp \left[
    i \int d^4 x d^8 \th \, \Theta_{R , \ga}^{(B)} (u; x, \th)
    \right]
    \, ,
    \label{2-46} \\
    \widehat{W}^{(0)} (x)
    & = &
    \exp \left[ -
    \int d \mu ( \widehat{P}_J ) \left(
    \frac{\del}{\del a_{J}^{(0)}} \otimes \frac{\del}{\del a_{-J}^{(0)}}
    \right) e^{ - i \widehat{P}_J \cdot ( x- y) }
    \right]_{y \rightarrow x \, ( x^0 > y^0 )}
    \nonumber \\
    & = &
    \exp \left[ -
    \int \frac{d^4  \widehat{P}_J }{(2 \pi )^4} \frac{i}{ \widehat{p}_J^2 - m^2 }
    \left(
    \frac{\del}{\del a_{J}^{(0)}} \otimes \frac{\del}{\del a_{-J}^{(0)}}
    \right) e^{ - i \widehat{P}_J \cdot ( x- y) }
    \right]_{y \rightarrow x \, ( x^0 > y^0 )}
    \label{2-47}
\eeqar
where in the calculation of $\widehat{W}^{(0)} (x)$
we take the limit $y \rightarrow x$, with the time ordering $x^0 > y^0$ being preserved.
An explicit derivation of the $x$-space UHV massive scalar tree amplitudes is
given by
\beqar
    &&
    \left.
    \frac{\del}{\del a_{1}^{(0)}} \otimes
    \frac{\del}{\del a_{2}^{(+)}} \otimes
    \frac{\del}{\del a_{3}^{(+)}} \otimes \cdots \otimes
    \frac{\del}{\del a_{n-1}^{(+)}} \otimes
    \frac{\del}{\del a_{n}^{(0)}}
    \, \F_{\rm UHV} \left[ a^{( \pm )c} , a^{(0)} \right]
    \right|_{a^{( \pm )c} = a^{(0)} = 0}
    \nonumber \\
    &=&
    \A_{\rm UHV}^{( \bar{\phi}_{1} g_{2}^{+} \cdots g_{n-1}^{+}  \phi_{n} )} (x)
    \label{2-48}
\eeqar
where $a^{( \pm )c}$ and $a^{(0)}$
represent generic source functions for gluons and massive scalars, respectively.

\section{S-matrix for massive fermion UHV amplitudes}

In the following sections, we consider the application
of the above results to massive fermion amplitudes.
In this section we focus on the massive fermion UHV amplitudes,
{\it i.e.}, the scattering amplitudes of an arbitrary number of
positive-helicity gluons and a pair of massive fermions.

\noindent
\underline{Notation}

We first clarify our definition of fermion operators.
In the spinor-helicity formalism fermions are represented
by two-component spinors. (For recent reviews of two-component
spinor techniques, see, {\it e.g.}, \cite{Dreiner:2008tw,Martin:2012us}.)
The ordinary Dirac spinor is defined by
\beq
    \Psi = \left(
             \begin{array}{c}
               \psi_{L}^{A} \\
               \psi_{R}^{\Ad} \\
             \end{array}
           \right)
           \label{3-1}
\eeq
where $\psi_L^A$ ($A = 1,2$) and $\psi_R^\Ad$ ($\Ad = 1,2$)
are the two-component Weyl spinors.
Each of these is analogous to the spinor momentum $u^A$.
Bearing in mind that its conjugate is
denoted as $\bu_\Ad = ( u^A )^*$ in our notation,
the conjugate of $\Psi$ can be defined as
\beq
    \bar{\Psi} = \Psi^\dag \ga^0
    = \left( \, (\psi_{L}^{A})^{*} ~~ (\psi_{R}^{\Ad})^{*} \, \right)
    \left(
      \begin{array}{cc}
        0 & 1 \\
        1 & 0 \\
      \end{array}
    \right)
    =
    \left( \, \bar{\psi}_{R \, A} ~~ \bar{\psi}_{L \, \Ad} \, \right)
    \label{3-2}
\eeq
where we use the chiral representation of the gamma matrix $\ga^0$.

For both massive and massless fermions, one can naturally assign helicity
$-\frac{1}{2}$ to $\psi_L$ and helicity $+ \frac{1}{2}$ to $\psi_R$, respectively.
(As we shall seen in a moment, one can make an off-shell continuation
of massless fermions such that the helicity of fermions is not altered; see
(\ref{3-18})-(\ref{3-21}).
Using such massive deformation, one can and should identify the `helicity'' of
massive fermions with the helicity of the corresponding massless fermions.)
In the helicity-spinor formalism, physical information of any
particle is encoded into its helicity and the numbering index.
In the framework of the holonomy formalism this is
represented by the physical operators $a_i^{(\pm )}( x, \th)$ in (\ref{2-19}).
With the help of the helicity operator (\ref{2-22}),
these physical operators are essentially described by
the $\N = 4$ supermultiplets in (\ref{2-21}).

Therefore, for {\it massless} fermions, we can
relate the Weyl spinors $\psi_L$ and $\psi_R$ to
the following operators:
\beqar
    \psi_{R i} & \longleftrightarrow & a_{Ri}^{\left( + \frac{1}{2} \right)} (\xi_i)
    \, = \, \xi_{i}^{\al}  \, a_{Ri \, \al}^{ \left( + \frac{1}{2} \right)} \, ,
    \label{3-3}
    \\
    \psi_{L i} & \longleftrightarrow &  a_{Li}^{\left( - \frac{1}{2} \right)} (\xi_i)
    \, = \,    \frac{1}{3!} \xi_{i}^{\al}\xi_{i}^{\bt}\xi_{i}^{\ga}
    \ep_{\al \bt \ga \del} \, a_{Li}^{ \left( - \frac{1}{2} \right)\, \del}
    \label{3-4}
\eeqar
where $i$ denotes the numbering index.
Similarly the conjugates of $\psi_L$ and $\psi_R$
have the following correspondence:
\beqar
    \bar{\psi}_{L i} & \longleftrightarrow & \bar{a}_{Li}^{\left( + \frac{1}{2} \right)} (\xi_i)
    \, = \, \xi_{i}^{\al}  \, \bar{a}_{Li \, \al}^{ \left( + \frac{1}{2} \right)} \, ,
    \label{3-5}
    \\
    \bar{\psi}_{R i} & \longleftrightarrow &  \bar{a}_{Ri}^{\left( - \frac{1}{2} \right)} (\xi_i)
    \, = \,    \frac{1}{3!} \xi_{i}^{\al}\xi_{i}^{\bt}\xi_{i}^{\ga}
    \ep_{\al \bt \ga \del} \,  \bar{a}_{Ri}^{ \left( - \frac{1}{2} \right) \, \del } \, .
    \label{3-6}
\eeqar
In the chiral-superspace representation, creation operators of these fermions
are then defined in the same way as the expression (\ref{2-19}), with $a_{i}^{(\pm )}$ being
respectively replaced by  $a_{L i}^{\left( - \frac{1}{2} \right)}$,
$a_{R i}^{\left( + \frac{1}{2} \right)}$,
$\bar{a}_{R i}^{\left( - \frac{1}{2} \right)}$ and
$\bar{a}_{L i}^{\left( + \frac{1}{2} \right)}$.
In order to incorporate fermions in the holonomy formalism,
basically we do not have to use the Weyl spinors but
the above creation operators. Notice that
these operators do not behave as spinors (which satisfy the
ordinary Dirac equation) but as mere Grassmann variables.

We can check the validity of our definitions (\ref{3-3})-(\ref{3-6})
for the massless fermions as follows.
From the study of fermionic extension of
the CSW rules, we find that the next-to-UHV (NUHV)
vertices, {\it i.e.}, vertices of an arbitrary number of
positive-helicity gluon plus one negative-helicity gluon
and a pair of fermions, can be calculated as
\cite{Ettle:2008ey,Schwinn:2008fm}:
\beqar
    V_{\rm MHV} (  \, g_i^- \,  g_j^- \, ) &=& \frac{(ij)^4}{(12)(23) \cdots (n-1 \, n)(n1)} \, ,
    \label{3-7} \\
    V_{\rm NUHV} ( \, \bar{\psi}_{R1} \, g_i^-  \, \psi_{R n} \, ) &=&
    \frac{(1i)^3 (i n) }{(12)(23) \cdots (n-1 \, n)(n1)} \, ,
    \label{3-8} \\
    V_{\rm NUHV} ( \, \bar{\psi}_{L1} \, g_i^-  \, \psi_{L n} \, ) &=&
    \frac{(1i) (i n)^3 }{(12)(23) \cdots (n-1 \, n)(n1)}
    \label{3-9}
\eeqar
where we include the MHV vertex for clarification of the notation;
here the positive-helicity gluons are implicit in the arguments
of the left-hand sides, and the color factors are omitted in
the right-hand sides.
Notation of fermions here is different from
the one found in the literature. For example, in
\cite{Schwinn:2008fm} light-cone
treatment of fermions is used and
they are denoted by $\bar{\psi}^\pm$, $\psi^\pm$;
these are related to our chiral notation by
$\psi^-  \Leftrightarrow \psi_L$,
$\psi^+  \Leftrightarrow \psi_R$,
$\bar{\psi}^-  \Leftrightarrow \bar{\psi}_R$ and
$\bar{\psi}^+  \Leftrightarrow \bar{\psi}_L$.
Carrying out the Grassmann integrals that are analogous to
(\ref{2-38}), we can easily check that the assignments (\ref{3-3})-(\ref{3-6})
indeed lead to the NUHV vertices (\ref{3-8}) and (\ref{3-9}) for massless fermions.

\noindent
\underline{Massive fermion UHV vertices and the choice of reference spinors}

We continue our discussion to include massive fermions.
The massive fermion UHV vertices, the fermionic analogs
of (\ref{2-35}), are given by \cite{Schwinn:2008fm}:
\beqar
    \widehat{V}_{\rm UHV} ( \, \bar{\psi}_{R 1} \, \psi_{R n} \, ) &=&
    \frac{m^2 (\eta 1) (1 n) }{(12)(23) \cdots (n-1 \, n)(n \eta )} \, ,
    \label{3-10} \\
    \widehat{V}_{\rm UHV} ( \, \bar{\psi}_{L 1} \, \psi_{L n} \, ) &=&
    \frac{m^2  (1 n) (n \eta ) }{(\eta 1 )(12)(23) \cdots (n-1 \, n)}\, ,
    \label{3-11} \\
    \widehat{V}_{\rm UHV} ( \, \bar{\psi}_{R 1} \, \psi_{L n} \, ) &=&
    \frac{m  (1 n)^3 }{(12)(23) \cdots (n-1 \, n) (n 1) }\, ,
    \label{3-12} \\
    \widehat{V}_{\rm UHV} ( \, \bar{\psi}_{L 1} \, \psi_{R n} \, ) &=& 0
    \label{3-13}
\eeqar
where $\eta$ denotes the {\it identical} reference spinor
for the pair of massive fermions labeled by the numbering indices $(1, n)$:
\beqar
    u_1^A & \longrightarrow &
    \widehat{u}_1^A \, = \,  u_1^A + \frac{m}{( u_1 \eta ) } \eta^A \, ,
    \label{3-14} \\
    u_n^A & \longrightarrow &
    \widehat{u}_n^A \, = \,  u_n^A + \frac{m}{( u_n \eta ) } \eta^A \, .
    \label{3-15}
\eeqar
In (\ref{3-10})-(\ref{3-13}) we follow the notations of (\ref{3-8}) and (\ref{3-9}).
Notice that $\widehat{V}_{\rm UHV}$'s are denoted with hats,
meaning that the fermions $\psi_{L/R}$'s and $\bar{\psi}_{L/R}$'s are massive.
From here on, we shall use the notation ($\bar{\psi}_{L1}$,
$\psi_{Ln}$, $\bar{\psi}_{R1}$, $\psi_{Rn}$) to express {\it massive} fermions.

The first two vertices (\ref{3-10}) and (\ref{3-11})
are explicitly dependent on the identical reference spinor $\eta$
and would lead to a set of arbitrary expressions, depending on the choice of $\eta$'s.
These $\eta$-dependent vertices are consequences of the parametrization
(\ref{3-14}) and (\ref{3-15}).
This parametrization is useful particularly in defining off-shell momentum transfers running
along propagators and indeed it has been utilized
in the derivation of the CSW rules \cite{Cachazo:2004kj}.
In defining exterior (instead of interior) off-shell momenta,
however, we can use alternative parametrization as well.

For example, in \cite{Rodrigo:2005eu,Ferrario:2006np}
Rodrigo and others parametrize a pair of massive fermions, utilizing
two {\it distinct} reference spinors.
Our choice of the reference spinors for a pair of massive scalars
in (\ref{2-29}) or (\ref{2-30})
is essentially the same as Rodrigo's.
(The difference is that their parametrization imposes
an additional condition
$\hat{p}^\mu_1 + \hat{p}^\mu_n = p^\mu_1 + p^\mu_n$
but ours does not; see appendix of \cite{Ferrario:2006np}
for detail of this point.)

The $\eta$-dependent vertices (\ref{3-10}) and (\ref{3-11})
vanish with suitable choices of $\eta$.
{\it In order to properly discuss forms of the massive fermion UHV vertices
and amplitudes in general,
it is therefore indispensable to use
$\eta$-independent expressions.
Otherwise we can not avoid
ambiguities arose from different choices of off-shell parametrizations
or reference spinors.
The vertices (\ref{3-12}) and (\ref{3-13}) are
the only such $\eta$-independent UHV vertices known in the literature
and, hence, we shall focus on these vertices in what follows,
keeping our choice of reference spinors for the pair of
massive fermions in the form of (\ref{2-29}),
rather than (\ref{3-14}) and (\ref{3-15}), as in the case
of the massive scalars.}

\noindent
\underline{Massive fermions in the holonomy formalism}

In the massive scalar case, the off-shell continuation of
Nair's superamplitude method is carried out by
the use of the $\xi \zt$-prescription (\ref{2-25}):
\beq
    a_{i}^{(0)} ( \xi_i ) ~ \longrightarrow ~
    a_{i}^{(0)} ( \xi_i , \zt_i ) ~ = ~
    \xi_{i}^{1} \xi_{i}^{2} \xi_{i}^{3} \zt_{i}^{4} \, a_{i}^{(0)}
    \label{3-16}
\eeq
where we have introduced the ``massive'' Grassmann variable
\beq
    \zt^\al_i \, = \, \th_{A}^{\al} \frac{m}{( u_i \eta_i ) } \eta_i^A
    \, .
    \label{3-17}
\eeq

Similarly, we can define the off-shell continuation
of the massless fermion operators (\ref{3-3})-(\ref{3-6})
such that it leads the $\eta$-independent massive UHV
vertices (\ref{3-12}), (\ref{3-13}).
We find that the fermionic off-shell continuation can be carried out by
\beqar
    a_{Ri}^{\left( + \frac{1}{2} \right)} (\xi_i) ~ \longrightarrow ~
    a_{Ri}^{\left( + \frac{1}{2} \right)} ( \xi_i ) & = &
    \xi_{i}^{\al}  \, a_{Ri \, \al}^{ \left( + \frac{1}{2} \right)} \, ,
    \label{3-18} \\
    a_{Li}^{\left( - \frac{1}{2} \right)} (\xi_i) ~ \longrightarrow ~
    a_{Li}^{\left( - \frac{1}{2} \right)} ( \xi_i ) & = &
    \frac{1}{3!} \ep_{\al \bt \ga \del} \xi_{i}^{\al}\xi_{i}^{\bt}\xi_{i}^{\ga}
    \, a_{Li}^{ \left( - \frac{1}{2} \right)\, \del} \, ,
    \label{3-19} \\
    \bar{a}_{Li}^{\left( + \frac{1}{2} \right)} (\xi_i) ~ \longrightarrow ~
    \bar{a}_{Li}^{\left( + \frac{1}{2} \right)} ( \xi_i , \zt_i ) & = &
    \frac{1}{3!} \ep_{\al \bt \ga \del} \xi_i^\al \xi_i^\bt \zt_i^\ga
    \, \bar{a}_{Li}^{(+ \frac{1}{2} ) \del} \, ,
    \label{3-20} \\
    \bar{a}_{Ri}^{\left( - \frac{1}{2} \right)} (\xi_i) ~ \longrightarrow ~
    \bar{a}_{Ri}^{\left( - \frac{1}{2} \right)} ( \xi_i , \zt_i ) & = &
    \frac{1}{4} \xi_i^1 \xi_i^2 \xi_i^3 \xi_i^4 \zt_i^\al
    \, \bar{a}_{Ri \, \al}^{(- \frac{1}{2} ) }
    \label{3-21}
\eeqar
where $i = 1, n$. Notice that
the off-shell continuation by means of the $\xi\zt$-prescription is made only
for the conjugate fermions $\bar{\psi}_{Li}$ and $\bar{\psi}_{Ri}$.
The unbar fermions $\psi_{Li}$ and $\psi_{Ri}$ remain on-shell.
This is consistent with the fact that the nontrivial UHV
vertex (\ref{3-12}) is proportional to $m$.
The rest of the UHV vertices vanish upon the Grassmann integrals
over the chiral supervariables $\th^\al_A$.
This can easily be seen by counting the by counting the number of Grassmann
variables in each pair. The numbers are given by
8, 4, 6 and 6
for $(  \bar{\psi}_{R 1} , \psi_{L n}  )$,
$( \bar{\psi}_{L 1} , \psi_{R n}  )$,
$( \bar{\psi}_{R 1} , \psi_{R n}  )$ and
$( \bar{\psi}_{L 1} , \psi_{L n}  )$,
respectively. Thus, upon the Grassmann integral over $d^8 \th$,
only the first pair survives.

Our choice (\ref{3-18})-(\ref{3-21}) is also consistent with
a recent study on recursion relations for massive fermion currents
\cite{Britto:2012qi} where a pair of massive fermions are
parametrized by an off-shell conjugate fermion and
an on-shell fermion.

The chiral superspace representation of the massive fermion operators
can then be expressed as
\beqar
    \bar{a}_{L1}^{\left( + \frac{1}{2} \right)}(x, \th)  & = &
    \left. \int d \mu ( \widehat{p}_1 ) ~ \bar{a}_{L1}^{\left( + \frac{1}{2} \right)} ( \xi_1 , \zt_1 )
    ~  e^{ i x_\mu \widehat{p}_1^\mu }
    \right|_{ \xi_{1}^{\al} = \th_{A}^{\al} u_1^A ,
    \, \zt_{1}^{\al} = \th_{A}^{\al} w_1^A  }
    \label{3-22} \\
    \bar{a}_{R1}^{\left( - \frac{1}{2} \right)}(x, \th)  & = &
    \left. \int d \mu ( \widehat{p}_1 ) ~ \bar{a}_{R1}^{\left( - \frac{1}{2} \right)} ( \xi_1 , \zt_1 )
    ~  e^{ i x_\mu \widehat{p}_1^\mu }
    \right|_{ \xi_{1}^{\al} = \th_{A}^{\al} u_1^A ,
    \, \zt_{1}^{\al} = \th_{A}^{\al} w_1^A  }
    \label{3-23} \\
    a_{L n}^{\left( - \frac{1}{2} \right)} (x, \th)  & = &
    \left. \int d \mu ( \widehat{p}_n ) ~ a_{L n}^{\left( - \frac{1}{2} \right)}  ( \xi_n )
    ~  e^{ i x_\mu \widehat{p}_n^\mu }
    \right|_{ \xi_{n}^{\al} = \th_{A}^{\al} u_n^A }
    \label{3-24} \\
    a_{R n}^{\left( + \frac{1}{2} \right)} (x, \th)  & = &
    \left. \int d \mu ( \widehat{p}_n ) ~ a_{R n}^{\left( + \frac{1}{2} \right)} ( \xi_n )
    ~  e^{ i x_\mu \widehat{p}_n^\mu }
    \right|_{ \xi_{n}^{\al} = \th_{A}^{\al} u_n^A }
    \label{3-25}
\eeqar
where we specify the numbering indices
that are relevant to the UHV vertices of our interest.
The off-shell momenta $\widehat{p}_1^\mu $ and $\widehat{p}_n^\mu$
are defined by (\ref{2-30}). As in (\ref{2-28}), $w_1^A$ is
given by
\beq
    w_1^A \, = \, \frac{m}{(u_1 \eta_1 )} \eta_1^A
    \, = \, \frac{m}{(u_1 u_n )} u_n^A
    \label{3-26}
\eeq
where our choice of reference spinors is reflected.

\noindent
\underline{The UHV rules and the massive fermion UHV amplitudes}

The basic ingredients of the UHV rules are the UHV vertices and the propagators.
In the CSW-type rules, which are originally formulated
in twistor space, the vertices correspond to a set of lines
in twistor space and they are connected by scalar propagators.
Application of the UHV rules to the massive fermion
UHV amplitudes is then straightforward because (a) we already know
all types of the fermionic UHV vertices and (b)
these vertices are connected by massive propagators as
in the case of massive scalar amplitudes.

The fermionic UHV vertex $\widehat{V}_{\rm UHV} ( \bar{\psi}_{1} \, \psi_{n} )$
should be expanded by possible pairs of fermions.
Under our choice of reference spinors, only a certain type remains nonzero, {\it i.e.},
\beqar
    \widehat{V}_{\rm UHV} ( \bar{\psi}_{1} \, \psi_{n} )
    & = &
    \widehat{V}_{\rm UHV} ( \bar{\psi}_{R1} \, \psi_{Ln} )
    +
    \widehat{V}_{\rm UHV} ( \bar{\psi}_{L1} \, \psi_{Rn} )
    +
    \widehat{V}_{\rm UHV} ( \bar{\psi}_{L1} \, \psi_{Ln} )
    +
    \widehat{V}_{\rm UHV} ( \bar{\psi}_{R1} \,\psi_{Rn} )
    \nonumber \\
    &=&
    \widehat{V}_{\rm UHV} ( \bar{\psi}_{R1} \, \psi_{Ln} )
    ~ = ~
    \frac{(1 n)}{m} \widehat{V}_{\rm UHV} ( \bar{\phi}_{1}  \phi_{n} )
    \label{3-27}
\eeqar
where in the last equation we use (\ref{2-35}) and (\ref{3-12}),
denoting the color-stripped massive scalar UHV vertices by
$\widehat{V}_{\rm UHV}( \bar{\phi}_1  \phi_n )$.
For the UHV vertices
other than $\widehat{V}_{\rm UHV} ( \bar{\psi}_{R1} \, \psi_{Ln} )$,
the number of Grassmann variables does not reach
the saturating number 8.
Thus these vertices vanish upon execution the Grassmann integrals
unless the pair of fermions
couple to other particles, either massive or massless, such that
the total number of Grassmann variables becomes 8.
We shall consider such interactions in another paper.

In application of the UHV rules, we can obtain
the massive fermion UHV amplitudes from the diagrams
in Figure \ref{fighol0601} by replacing $(\bar{\phi}, \phi)$ with
$(\bar{\psi}_R, \psi_L )$.
Analytically the massive fermion UHV amplitudes
can be written down in the form of (\ref{2-42}):
\beq
    \widehat{C}_{\rm UHV}^{ ( \bar{\psi}_{1}  \psi_{n} ) } ( u )
    \, = \,
    \widehat{C}_{\rm UHV}^{ ( \bar{\psi}_{1} g_{2}^{+} g_{3}^{+}
    \cdots g_{n-1}^{+} \psi_{n} ) } ( u )
    \, = \,
    \frac{ - m (n1)^2 }{(12)(23)\cdots (n1)}
    \widehat{(n1)}_{\bar{\psi}  \psi}
    \label{3-28}
\eeq
where
\beqar
    \widehat{(n1)}_{\bar{\psi}  \psi}
    &=&
    (n1) \, + \, \sum_{j =2}^{n-2}
    \frac{(J1)^2}{(jJ)}
    \frac{- m (j \, j+1)}{ \widehat{P}_J^2 - m^2}
    \frac{(nJ)^2}{(J\, j+1) (n1)}
    \nonumber \\
    &&
    + \,
    \sum_{2 \le i < j \le n-1 }
    \frac{(I1)^2}{(iI)}
    \frac{ - m (i \, i+1)}{ \widehat{P}_I^2 - m^2}
    \frac{(JI)^2}{(I \, i+1)(j J)}
    \frac{ - m (j \, j+1)}{ \widehat{P}_J^2 - m^2}
    \frac{(nJ)^2}{(J \, j+1) (n1)}
    \nonumber \\
    &&
    + \, \sum_{2 \le i < j < k \le n-1 }  \, \Biggl[ \,
    \frac{(I1)^2}{(iI)}
    \frac{ - m (i \, i+1)}{ \widehat{P}_I^2 - m^2}
    \frac{(JI)^2}{(I \, i+1)(j J)}
    \frac{ - m (j \, j+1)}{ \widehat{P}_J^2 - m^2}
    \nonumber \\
    &&
    ~~~~~~~~~~~~~~~~~~~~~~~~~~~
    \times \,
    \frac{(KJ)^2}{(J \, j+1)(k K)}
    \frac{ - m (k \, k+1)}{ \widehat{P}_K^2 - m^2}
    \frac{(nK)}{(K \, k+1) (n1) }
    \,   \Biggr]
    \nonumber \\
    &&
    + \, \cdots \,
    \nonumber \\
    &=&
    -
    \left( 1 \left|
    \prod_{j = 2}^{n-2} \left[
    1 + \frac{ |J )(1J) }{(j J) }
    \frac{  m ( j \, j+1 ) }{\widehat{P}_J^2 - m^2}
    \frac{ (J n)(J | }{(J \, j+1 ) }
    \right]
    \right| n \right) \, .
    \label{3-29}
\eeqar
The full amplitudes in a form holomorphic to the spinor momenta
are given, as in (\ref{2-41}), by
\beq
    \widehat{A}_{\rm UHV}^{( \bar{\psi}_1  \psi_n )} (u)
    \, = \,
    \Tr ( t^{c_{2}} t^{c_{3}} \cdots t^{c_{n-1}}) \,
    \widehat{C}_{\rm UHV}^{(  \bar{\psi}_1 \psi_n  )} (u)
    \, .
    \label{3-30}
\eeq
We should emphasize again that
we do not take a sum over permutations of gluons
in the above expression; the sum is already taken
care of in the definition of the massive holonomy operator.
To be more precise, the sum is already included
in the computation of the braid trace for the
massive holonomy operator; see (\ref{2-17}) and (\ref{2-18})
for details.

We should comment on the relation between
the massive UHV amplitudes of scalars and fermions.
In the literature it is often shown that these
are proportional to each other as in
the case of the UHV vertices (\ref{3-27}),
{\it i.e.},
$\widehat{C}_{\rm UHV}^{( \bar{\psi}_{1}  \psi_{n} )} (u)
= \frac{(1 n)}{m} \widehat{C}_{\rm UHV}^{( \bar{\phi}_{1}  \phi_{n} )}(u)$.
This relation is derived from the supersymmetric Ward identities
\cite{Schwinn:2008fm}; see also \cite{Schwinn:2006ca,Huang:2012gs,Boels:2011zz}.
We consider, however, that this relation is
contradictory to the UHV rules because in the
UHV rules the massive UHV amplitudes
are constructed by iterative use of the UHV vertices, see Figure \ref{fighol0601}
or the analytic expression (\ref{3-29}).
Therefore the above proportional relation
holds only for the 1 UHV irreducible (1UI) amplitudes
or the UHV vertices in our framework.
This interpretation is natural and would improve our understanding
of the massive UHV amplitudes since the notion
of 1UI amplitudes and its roles in construction of massive UHV amplitudes
have been generally ignored so far in the literature.

\noindent
\underline{An S-matrix functional for the massive fermion UHV amplitudes}

What we have shown so far is that based on the forms of the massive fermion
UHV vertices (\ref{3-10})-(\ref{3-13}), one can apply the UHV rules to construct
the massive fermion UHV amplitudes in much the same way as the case of
massive scalar amplitudes.
Motivated by the results in (\ref{2-46})-(\ref{2-48}),
we now formulate field theoretic derivation of the massive fermion UHV amplitudes.
Since the massive holonomy operator (\ref{2-14}) does not contain
fermion operators (\ref{3-18})-(\ref{3-21}), for this purpose, we need
to incorporate the operators of massive fermions, rather than
those of massive scalars $( a_{1}^{(0)} , a_{n}^{(0)})$, into the
massive holonomy operator. As in (\ref{3-27}), this
can be done by taking a sum over possible pairs of fermions, {\it i.e.},
\beqar
    &&
    \Theta_{R, \ga}^{(B)_{\bar{\psi}\psi} } (u)
    \nonumber \\
    &=& \exp \Biggl[ \,
    \sum_{r \ge 3}
    \sum_{ ( h_2 , h_3 , \cdots , h_{r-1} ) }
    g^{r-2} ( -1 )^{ h_2 h_3 \cdots h_{r-1} }
    \,
    \Tr \left(
    t^{c_2} t^{c_3} \cdots t^{c_{r-1}}
    \right)
    \frac{
    a_{2}^{(h_2)c_2} \otimes \cdots \otimes a_{r-1}^{(h_{r-1})c_{r-1}}
    }{
    (12)(23) \cdots (r-1 \, r)( r 1)
    }
    \nonumber \\
    &&
    ~~~~~~ \otimes \,
    \left(
    a_{Lr}^{\left( - \frac{1}{2} \right)} \otimes \bar{a}_{R1}^{\left( - \frac{1}{2} \right)}
    +
    a_{Rr}^{\left( + \frac{1}{2} \right)} \otimes \bar{a}_{L1}^{\left( + \frac{1}{2} \right)}
    +
    a_{Lr}^{\left( - \frac{1}{2} \right)} \otimes \bar{a}_{L1}^{\left( + \frac{1}{2} \right)}
    +
    a_{Rr}^{\left( + \frac{1}{2} \right)} \otimes \bar{a}_{R1}^{\left( - \frac{1}{2} \right)}
    \right) \,
    \Biggr]
    \label{3-31}
\eeqar
where $h_{i} = \pm = \pm 1$ ($i=2,3, \cdots, r-1$) denotes
the helicity of the $i$-th gluon.
We obtain the above expression based on the UHV rules, which means that
we do not need a knowledge of algebra for the fermionic operators.
This is related to the fact that in the holonomy formalism the fermionic
operators arise upon supersymmetrization of the gluon operators; see (\ref{2-21}).

Supersymmetrization of $\Theta_{R , \ga}^{(B)_{\bar{\psi} \psi} } (u)$
can be carried out by replacing the gluon and fermion operators with
(\ref{2-19}) and (\ref{3-22})-(\ref{3-25}), respectively.
As in the massive scalar case, we denote the supersymmetric
holonomy operator by
$\Theta_{R , \ga}^{(B)_{\bar{\psi} \psi} } (u; x, \th)$.
The generating functional for the massive fermion UHV vertices is then
expressed as
\beq
    \F_{\rm UHV}^{\rm (vertex)} \left[ a^{( \pm )c} ,
    \bar{a}_{L/R}^{\left( \pm \frac{1}{2} \right)} ,
    a_{L/R}^{\left( \mp \frac{1}{2} \right)}
    \right]
    \, = \,
    \exp \left[
    i \int d^4 x d^8 \th ~ \Theta_{R , \ga}^{(B)_{\bar{\psi} \psi} } (u; x, \th)
    \right] \, .
    \label{3-32}
\eeq
This expression provides a fermionic analog of (\ref{2-31}).

Similarly, in analogy to (\ref{2-46}) and (\ref{2-47}),
we can straightforwardly obtain an S-matrix functional for
the full massive fermion UHV amplitudes
by use of $\Theta_{R , \ga}^{(B)_{\bar{\psi} \psi}} (u; x, \th)$:
\beqar
    \F_{\rm UHV}  \left[ a^{(h)c}, \bar{a}_{L/R}^{\left( \pm \frac{1}{2} \right)} ,
    a_{L/R}^{\left( \mp \frac{1}{2} \right)}  \right]
    & = &
    \widehat{W}^{(B)}_{ \bar{R}L } (x)
    \widehat{W}^{(B)}_{ \bar{L}R } (x)
    \widehat{W}^{(B)}_{ \bar{L}L } (x)
    \widehat{W}^{(B)}_{ \bar{R}R } (x)\,
    \nonumber \\
    &&
    ~~~ \times \,
    \exp \left[
    i \int d^4 x d^8 \th ~ \Theta_{R , \ga}^{(B)_{\bar{\psi} \psi} } (u; x, \th)
    \right]
    \label{3-33}
\eeqar
where
\beqar
    \widehat{W}^{(B)}_{ \bar{R}L } (x)
    & = &
    \exp \left[ -
    \int d \mu ( \widehat{P}_J ) \left(
    \frac{\del}{\del \bar{a}_{RJ \, \al}^{\left( - \frac{1}{2} \right)} } \otimes
    \frac{\del}{\del a_{LJ}^{\left( - \frac{1}{2} \right) \al}}
    \right) e^{ - i \widehat{P}_J \cdot ( x- y) }
    \right]_{y \rightarrow x \, ( x^0 > y^0 )} \, ,
    \label{3-34} \\
    \widehat{W}^{(B)}_{ \bar{L}R } (x)
    & = &
    \exp \left[ -
    \int d \mu ( \widehat{P}_J ) \left(
    \frac{\del}{\del \bar{a}_{LJ}^{\left( + \frac{1}{2} \right) \al} } \otimes
    \frac{\del}{\del a_{RJ \, \al}^{\left( + \frac{1}{2} \right)}}
    \right) e^{ - i \widehat{P}_J \cdot ( x- y) }
    \right]_{y \rightarrow x \, ( x^0 > y^0 )} \, ,
    \label{3-35} \\
    \widehat{W}^{(B)}_{ \bar{L}L } (x)
    & = &
    \exp \left[ -
    \int d \mu ( \widehat{P}_J ) \left(
    \frac{\del}{\del \bar{a}_{LJ}^{\left( + \frac{1}{2} \right) \al} } \otimes
    \frac{\del}{\del a_{LJ}^{\left( - \frac{1}{2} \right) \al}}
    \right) e^{ - i \widehat{P}_J \cdot ( x- y) }
    \right]_{y \rightarrow x \, ( x^0 > y^0 )} \, ,
    \label{3-36} \\
    \widehat{W}^{(B)}_{ \bar{R}R } (x)
    & = &
    \exp \left[ -
    \int d \mu ( \widehat{P}_J ) \left(
    \frac{\del}{\del \bar{a}_{RJ \, \al}^{\left( - \frac{1}{2} \right)} } \otimes
    \frac{\del}{\del a_{RJ \, \al}^{\left( + \frac{1}{2} \right)}}
    \right) e^{ - i \widehat{P}_J \cdot ( x- y) }
    \right]_{y \rightarrow x \, ( x^0 > y^0 )}
    \label{3-37}
\eeqar
where off-shell momentum transfer $\widehat{P}_J^\mu$ is defined by (\ref{2-44}).
The Lorentz invariant measure $d \mu ( \widehat{P}_J )$ is given by
\beq
    d \mu ( \widehat{P}_J ) \, = \, \frac{d^3 \widehat{P}_J }{(2 \pi)^3}
    \frac{1}{2 \sqrt{|\vec{\widehat{P}}_{J}|^{2} + m^2 }}
    \label{3-38}
\eeq
where $m$ denotes the mass of the fermion.
As before, in the calculation of (\ref{3-34})-(\ref{3-37}) we take
the limit $y \rightarrow x$, while keeping the time ordering $x^0 > y^0$.

The massive fermion UHV amplitudes in the $x$-space
representation are then explicitly derived as
\beqar
    &&
    \frac{\del}{\del a_{2}^{(+)}} \otimes
    \frac{\del}{\del a_{3}^{(+)}} \otimes \cdots \otimes
    \frac{\del}{\del a_{n-1}^{(+)}} \otimes
    \left[
    \frac{\del}{\del a_{Ln}^{\left( - \frac{1}{2} \right)} }
    \otimes \frac{\del}{\del \bar{a}_{R1}^{\left( - \frac{1}{2} \right)} }
    +
    \frac{\del}{\del a_{Rn}^{\left( + \frac{1}{2} \right)} }
    \otimes \frac{\del}{\del \bar{a}_{L1}^{\left( + \frac{1}{2} \right)} }
    \right.
    \nonumber \\
    && ~
    \left.
    \left.
    +
    \frac{\del}{\del a_{Ln}^{\left( - \frac{1}{2} \right)} }
    \otimes \frac{\del}{\del \bar{a}_{L1}^{\left( + \frac{1}{2} \right)} }
    +
    \frac{\del}{\del a_{Rn}^{\left( + \frac{1}{2} \right)} }
    \otimes \frac{\del}{\del \bar{a}_{R1}^{\left( - \frac{1}{2} \right)} }
    \right]
    \, \F_{\rm UHV} \left[ a^{( \pm )c} , a^{(0)} \right]
    \right|_{a^{( \pm )c} = {\bar a}^{(\pm \hf)} = a^{(\mp \hf)} = 0}
    \nonumber \\
    &=&
    \A_{\rm UHV}^{( \bar{\psi}_{1} g_{2}^{+} \cdots g_{n-1}^{+}  \psi_{n} )} (x)
    \label{3-39}
\eeqar
where $a^{( \pm )c}$ and $( {\bar a}^{(\pm \hf)} , a^{(\mp \hf)} )$
represent generic source functions for gluons and massive fermions, respectively.

\noindent
\underline{Dimensional analysis and ``exclusion rules'' for massive fermion amplitudes}

Lastly, but most importantly, we consider the $m$ dependence of the massive fermion UHV
amplitudes $\widehat{A}_{\rm UHV}^{( \bar{\psi}_1  \psi_n )} (u)$.
In the massive scalar case, the UHV amplitudes are
proportional to the squared mass,
$\widehat{A}_{\rm UHV}^{( \bar{\phi}_1  \phi_n )} (u) \sim m^2$.
Owing to the relation (\ref{3-27}) and the form of the massive propagator,
this proportionality does not hold for
$\widehat{A}_{\rm UHV}^{( \bar{\psi}_1  \psi_n )} (u)$.
Let $s$ be the number of 1 UHV irreducible (1UI)
diagrams in the expansion formula of Figure \ref{fighol0601}.
Then we find that
$\widehat{A}_{\rm UHV}^{( \bar{\psi}_1  \psi_n )} (u)$
is expanded by $m^{s - 2(s -1)} = m^{-s +2}$ ($s = 1,2, \cdots , n-2$).
This shows that in large $m$ limits we can approximate
$\widehat{A}_{\rm UHV}^{( \bar{\psi}_1  \psi_n )} (u)$
to the 1UI subamplitudes or the UHV vertices
$\widehat{V}_{\rm UHV}^{( \bar{\psi}_1  \psi_n )} (u)$.

The fact that
$\widehat{A}_{\rm UHV}^{( \bar{\psi}_1  \psi_n )} (u)$
is expanded by $m^{-s +2}$ ($s = 1,2, \cdots , n-2$) is
unnatural from a perspective that
the fermion-mass dependence is
irrelevant to the number of gluons involving the scattering processes.
The only way to circumvent this problem is to fix
the total number of scattering particles by
\beq
    n  \, = \, 3 \, .
    \label{3-40}
\eeq
Naively, this constraint leads to the relation
\beq
    \widehat{A}_{\rm UHV}^{( \bar{\psi}_1  g_2^+ \psi_3 )} (u)
    \, = \, \Tr ( t^{c_2} ) \frac{- m (31)^2 }{(12)(23)}
    \label{3-41}
\eeq
where the color factor $\Tr ( t^{c_2} )$ vanishes unless the gluon propagates along
an internal line.  As in the case of the MHV rules, internal gluons have the
$U(1)$ color factor. But, for external gluons, we have the $SU(N)$ color factor and
$\widehat{A}_{\rm UHV}^{( \bar{\psi}_1  g_2^+ \psi_3 )} (u)$ vanishes.
To be explicit, $\Tr ( t^{c_2} )$ is given by
\beq
    \Tr ( t^{c_2} ) = \left\{
    \begin{array}{cl}
      0 & ~~ \mbox{for $t^{c_2} \in \underline{SU(N)}$ } \\
      \sqrt{ \frac{N}{2} } & ~~ \mbox{for $t^{c_2} \in \underline{U(1)}$ }
    \end{array}
    \right.
    \label{3-42}
\eeq
where $\underline{SU(N)}$ and $\underline{U(1)}$
denote the algebras of $SU(N)$ and $U(1)$ groups, respectively.
We here use the usual normalization $\Tr ( t^c t^{c'}) = \frac{\del^{c c'}}{2}$.
The three-point interactions (\ref{3-41}) vanish unless the involving
gluon is a virtual gluon which appears only in internal propagators.
Intriguingly, this feature is in accord with the pictures of the
quark-gluon interactions in QCD.

In the existing literature there have been no considerations
on the number of involving particles in the massive UHV amplitudes.
This is mainly because the notion of the 1 UHV irreducible (1UI) diagram
in Figure \ref{fighol0601} is not well-recognized in the literature.
In our formulation the 1UI diagram corresponds to the massive UHV {\it vertex}
but it is often interpreted as the massive UHV {\it amplitude};
it is so particularly in the case of massive fermions.
Once this point is clarified, one can utilize the relation (\ref{3-27})
to derive the explicit form of the massive fermion UHV amplitudes (\ref{3-28}).
We can then straightforwardly carry out mass-dimension analysis of
the UHV amplitudes. For scalar amplitudes a pair of massive scalars
can couple to an arbitrary number of gluons. This is essentially due
to the fact that the massive scalar UHV vertex (\ref{2-35}) is proportional
to $m^2$. Since the massive propagator is proportional to $m^{-2}$, one
can insert as many propagators as possible into the massive scalar
UHV amplitudes as far as the mass dimension is concerned.
On the other hand, the massive fermion UHV vertex (\ref{3-27}) is proportional
to $m$. Thus insertion of massive propagators is mass-dimensionally prohibited.
This leads to the constraint (\ref{3-40}) that a pair of massive fermions
should couple to a single (internal) gluon.

The constraint $n=3$ is crucial when we consider the generalization
of the massive fermion UHV amplitudes to the non-UHV amplitudes.
For example, it means that all the massive fermion amplitudes
are classified by the UHV subamplitude and the next-to-UHV (NUHV) subamplitude,
the latter containing a single negative-helicity gluon.
In this sense, we can interpret the condition $n=3$ for a sort of
exclusion rules for the massive fermion amplitudes.
This principle is a consequence of (a)
the use of massive scalar propagators for the construction of
massive fermion amplitudes in the holonomy formalism
and (b) the mass-dimension analysis of the massive fermion UHV amplitudes.

In the following section, for clarification of our argument, we
first consider the massive fermion NUHV amplitudes
for arbitrary $n$ and then
study the consequence of the constraint $n = 3$ afterwards.

\section{Massive fermion NUHV amplitudes}

\noindent
\underline{Massive fermion NUHV vertices}

For the massive scalar amplitudes
there are no vertices other than the massive scalar UHV vertices
that contribute to the UHV rules.
For the massive fermion amplitudes, however, due to our assignments of
Grassmann variables in (\ref{3-3})-(\ref{3-4}),
there does exists another type of nonzero vertices
that contributes to the UHV rules.
This is given by the following NUHV vertices
\beq
    \widehat{V}_{\rm NUHV} (  \bar{\psi}_{L 1} \, g_a^- \, \psi_{R n}  )
    \,= \,
    \frac{m}{(1n)} \frac{ (1 \, a)^2 (a \, n)^2 }{(12)(23) \cdots (n-1 \, n) (n 1) }
    \label{4-1}
\eeq
($2 \le a \le n-1$)
where we choose the reference spinor $\eta_1 = u_n$ as before
for the operator (\ref{3-20}) of the
off-shell fermion $\bar{\psi}_{L 1}$.
Notice that one can easily check that there are no other
non-vanishing massive fermion vertices, hence,
(\ref{3-12}) and (\ref{4-1}) are the only vertices that
contribute to the massive fermion amplitudes in general.

\noindent
\underline{Massive fermion NUHV amplitudes}

\begin{figure} [htbp]
\begin{center}
\includegraphics[width=150mm]{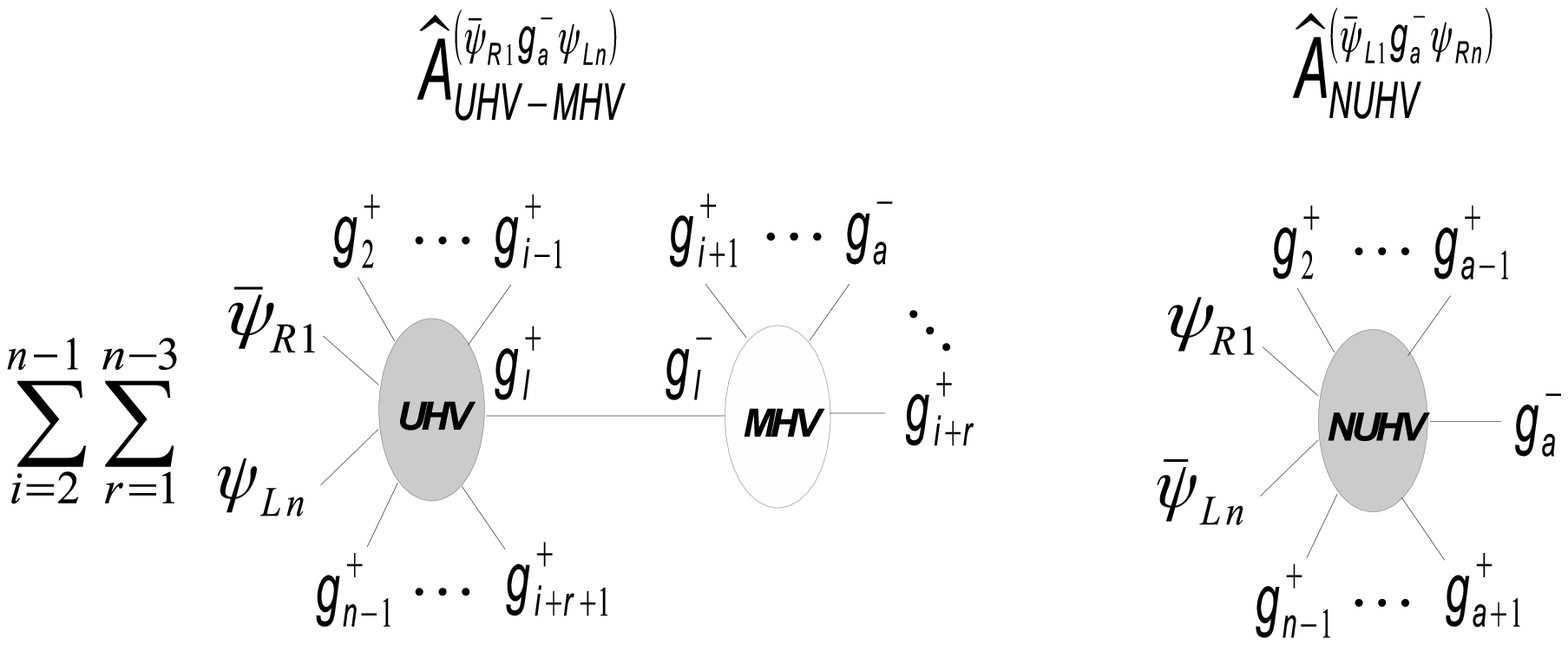}
\caption{Diagrams contributing to the massive fermion NUHV amplitudes}
\label{fighol0602}
\end{center}
\end{figure}

The contribution (\ref{4-1}) is one of the major differences
between the UHV rules of massive scalars and fermions.
We here study how it reflects in the construction of the
massive fermion NUHV amplitudes.
Diagrams contributing to the massive fermion NUHV amplitudes
$\widehat{A}_{\rm NUHV}^{( \bar{\psi}_{1} g_{a}^{-}  \psi_{n} )} (u)$
are shown in Figure \ref{fighol0602}.

There are two types of contributions to
$\widehat{A}_{\rm NUHV}^{( \bar{\psi}_{1} g_{a}^{-}  \psi_{ n} )} (u)$:
\beq
    \widehat{A}_{\rm NUHV}^{( \bar{\psi}_{1} g_{a}^{-}  \psi_{ n} )} (u)
    ~ = ~
    \widehat{A}_{\rm UHV-MHV}^{( \bar{\psi}_{R 1} g_{a}^{-}  \psi_{L n} )} (u)
    ~+~
    \widehat{A}_{\rm NUHV}^{( \bar{\psi}_{L 1} g_{a}^{-}  \psi_{R n} )} (u)
    \, .
    \label{4-2}
\eeq
One includes the gluonic MHV amplitudes and the other does not.
From the results of
$\widehat{A}_{\rm NUHV}^{( \bar{\phi}_{1} g_{a}^{-}  \phi_{ n} )} (u)$
in \cite{Abe:2012en}, we can write down the former contribution
$\widehat{A}_{\rm UHV-MHV}^{( \bar{\psi}_{R 1} g_{a}^{-}  \psi_{L n} )} (u)$
as
\beqar
    &&
    \widehat{A}_{\rm UHV-MHV}^{( \bar{\psi}_{R 1} g_{a}^{-}  \psi_{L n} )} (u)
    \nonumber \\
    & = &
    \left.
    \sum_{i = 2}^{n-1} \sum_{r = 1}^{n-3}
    \widehat{A}^{( (i+r+1)_{+} \cdots \psi_{L n} \bar{\psi}_{R 1}
    \cdots (i-1)_{+} l_{+} )}_{\rm UHV} (u)
    \, \frac{1}{q_{i \, i+r}^2} \,
    \widehat{A}^{( l_{-} \, (i)_{+} \cdots a_{-} \cdots (i+r)_{+} )}_{\rm MHV} (u)
    \right|_{u_l = u_{i \, i+r}}
    \nonumber \\
    &=&
    \sum_{i = 2}^{n-1}
    \sum_{r = 1}^{n - 3 }
    \sum_{ \si \in \S_{r+1}} \!\!\!
    \Tr ( t^{\si_{i}} \cdots t^{\si_{i+r}}  \,
    t^{i+r+1} \cdots t^{i-1} )
    \nonumber \\
    &&
    \left.
    \widehat{C}^{( (i+r+1)_{+} \cdots  \psi_{L n} \bar{\psi}_{R 1}
    \cdots (i-1)_{+} l_{+} )}_{\rm UHV} (u)
    \, \frac{1}{q_{\si_{i} \si_{i+r}}^2} \,
    \widehat{C}^{( l_{-} \, (i)_{+} \cdots a_{-} \cdots (i+r)_{+} )}_{\rm MHV} (u; \si)
    \right|_{u_l = u_{\si_{i}  \si_{i+r} }}
    \label{4-3}
\eeqar
where we consider the numbering indices in modulo $n$.
The $\widehat{C}$'s are expressed as
\beqar
    \widehat{C}^{( (i+r+1)_{+} \cdots \psi_{L n} \bar{\psi}_{R 1}
    \cdots (i-1)_{+} l_{+} )}_{\rm UHV} (u)
    &=&
    \frac{ -m (n1)^2  \widehat{(n1)}_{\bar{\psi} \psi } }{
    (i+r+1 ~ i+r+2) \cdots (i-1 ~ l)(l \, i+r+1 )
    } \, ,
    \label{4-4}\\
    \widehat{C}^{( l_{-} \, (i)_{+} \cdots a_{-} \cdots (i+r)_{+} )}_{\rm MHV} (u; \si)
    &=&
    \frac{(l a)^4 }{
    (l \, \si_i )( \si_i \, \si_{i+1} ) (\si_{i+1} \, \si_{i+ 2} ) \cdots (\si_{i+r} \, l)
    }
    \label{4-5}
\eeqar
where the off-shell momentum transfer
$q_{i \, i+r}^{\mu}$ is defined in terms of the
reference null-vector $\eta_{i \, i+r}^{\mu}$
and the associated
null momentum $p_{i \, i+r}^{\mu}$:
\beqar
    q_{i \, i+r}^{\mu}
    & = &
    p_{i}^{\mu} + p_{i+1}^{\mu} + \cdots + p_{i+r}^{\mu}
    ~ \equiv ~
    p_{i \, i+r}^{\mu} + W \eta_{i \, i+r}^{\mu}
    \, ,
    \label{4-6} \\
    p_{i \, i+r}^{A \Ad} & = & u_{i \, i+r}^A \bu_{i \, i+r}^\Ad
    \, \equiv \, u_{l}^A \bu_{l}^\Ad
    \, .
    \label{4-7}
\eeqar
In (\ref{4-7}) we use the two-component notation of the
null vector $p_{i \, i+r}^{\mu}$.
In (\ref{4-3}) and (\ref{4-5}) $\si$
denotes the transposition of the numbering indices for gluons:
\beq
    \si=\left(%
    \begin{array}{l}
      i ~~\, i+1 ~ \cdots ~ i+r  \\
      \si_{i} ~~ \si_{i+1} ~ \cdots ~ \si_{i+r} \\
    \end{array}%
    \right)  .
    \label{4-8}
\eeq

The other contribution to
$\widehat{A}_{\rm NUHV}^{( \bar{\psi}_{1} g_{a}^{-}  \psi_{ n} )} (u)$ is
essentially
given by the massive fermion NUHV vertices:
\beq
    \widehat{A}_{\rm NUHV}^{( \bar{\psi}_{L 1} g_{a}^{-}  \psi_{R n} )} (u)
    ~=~
    \Tr ( t^{c_{2}} t^{c_{3}} \cdots t^{c_{n-1}}) \,
    \frac{m}{(1n)} \frac{ (1 \, a)^2 (a \, n)^2 }{(12)(23) \cdots (n-1 \, n) (n 1) }
    \, .
    \label{4-9}
\eeq
This expression is free of the reference spinors.

\noindent
\underline{Imposition of the $n=3$ condition}

Having written down the massive fermion NUHV amplitudes
for arbitrary $n$, we now impose the condition $n=3$ of (\ref{3-40})
on (\ref{4-2}). Since the MHV vertex needs more than two legs,
the first term in (\ref{4-2}) vanishes for $n =3$, {\it i.e.},
\beqar
    \widehat{A}_{\rm NUHV}^{( \bar{\psi}_{1} g_{2}^{-}  \psi_{3} )} (u)
    \, = \,
    \widehat{A}_{\rm NUHV}^{( \bar{\psi}_{L 1} g_{2}^{-}  \psi_{R 3} )} (u)
    \, = \,
    \frac{ - m (12)(23) }{ (31)^2 }
    \label{4-10}
\eeqar
where we omit the color factor.
As discussed in (\ref{3-41}) and (\ref{3-42}),
the gluon labeled by $g_2^-$ is interpreted as a virtual gluon
otherwise the above NUHV amplitude becomes zero. In practice, however,
one can connect it with the three-point
UHV subamplitude (\ref{3-41}), in which case the color factor becomes nonzero
(regardless the choice of the color for each gluon)
due to the normalization $\Tr ( t^c t^{c'}) = \frac{\del^{c c'}}{2}$.
The color-stripped UHV amplitude can similarly be expressed as
\beqar
    \widehat{A}_{\rm UHV}^{( \bar{\psi}_{1} g_{2}^{+}  \psi_{3} )} (u)
    \, = \,
    \widehat{A}_{\rm UHV}^{( \bar{\psi}_{R 1} g_{2}^{+}  \psi_{L 3} )} (u)
    \, = \,
    \frac{- m (31)^2 }{(12)(23)}
    \, .
    \label{4-11}
\eeqar

In general non-UHV amplitudes are labeled by N$^k$UHV amplitudes,
specified by the number of negative-helicity gluons, $k = 1,2, \cdots n-3$.
The massive fermion N$^k$UHV amplitudes then refer to
the scattering amplitudes of a pair of massive fermions
and $k$ negative-helicity gluons and ($n-k-2$) positive-helicity gluons.
Therefore, upon the imposition of $n =3$, there exist no
N$^k$UHV amplitudes for $k \ge 2$.

\noindent
\underline{An S-matrix functional for (\ref{4-10}) and (\ref{4-11})}

To summarize, we find that the
interactions among gluons and a pair of massive fermions
are given by
(\ref{4-10}) and (\ref{4-11}).
This result is qualitatively different from
the massive scalar amplitudes where
we can incorporate an arbitrary number of gluons.
The difference arises from the
the mass dimension analysis
which leads to the condition $n=3$.

Notice that, apart from the $n=3$ constraint, (\ref{4-10}) and (\ref{4-11})
are obtained by application of the UHV rules.
Since the UHV rules are automatically realized by
defining an S-matrix functional, it is still worth
deriving the S-matrix functional for
(\ref{4-10}) and (\ref{4-11}).
Since the massive fermion NUHV vertices (\ref{4-1})
can be obtained from the generating functional
for the UHV vertices (\ref{3-32}) alone, we find
that the amplitudes in (\ref{4-10}) are
derived from (\ref{3-33}) as well.
Therefore, together with (\ref{3-34})-(\ref{3-37}),
the S-matrix functional for (\ref{4-10}) and (\ref{4-11})
is given by (\ref{3-33}).

\section{Concluding remarks}

In this paper we consider incorporation of massive fermions
into the holonomy formalism, following the recent study \cite{Abe:2012en}
on the massive scalar amplitudes.
The upshot of this paper is that
the interactions among gluons and a pair of fermions
are reduced to the two types of subamplitudes represented by
(\ref{4-10}) and (\ref{4-11}).
This result is simple but there have been
several key steps toward it. These steps can be itemized as follows:
\begin{enumerate}
  \item We clarify our notation of fermions in the spinor-helicity formalism.
  \item We carry out off-shell continuation of Nair's  superamplitude method for fermions
  by use of what we call the $\xi \zt$-prescription in (\ref{3-16})-(\ref{3-21}). This
  set of parametrizations leads to the previously known massive fermion UHV vertices \cite{Schwinn:2008fm}.
  \item We apply the UHV rules, {\it i.e.}, the massive extension of the CSW rules, to
  construct massive fermion UHV amplitudes.
  \item In the CSW-type rules, vertices correspond to a set of lines in twistor space
  and they are connected by scalar propagators.
  In the UHV rules massive fermion amplitudes are therefore obtained by connecting
  the massive fermion UHV vertices in terms of massive scalar propagators.
  \item We formulate an S-matrix functional for the massive fermion UHV amplitudes
  that realizes the UHV rules in a functional method.
  \item Analysis of mass dimension on the massive fermion UHV amplitudes implies
  that the number of gluons involving the UHV amplitudes should be one.
  \item Interactions among gluons and a pair of massive fermions are then
  decomposed into the two types of subamplitudes
  (\ref{4-10}) and (\ref{4-11}).
\end{enumerate}

The final result 7 means that interactions among gluons and massive
fermions should be decomposed into three-point massive fermions
vertices. This is consistent with QCD in terms of the gluon-quark interactions.
It is intriguing that this result is derived purely from the mass-dimension analysis.
In the framework of the holonomy formalism, the result 7 also
illustrates a qualitative difference from the massive scalar amplitudes.
In the massive scalar amplitudes there are no restrictions on the number
of involving gluons. In this sense, we can interpret the result 6 as a sort of
``exclusion rules'' for the massive fermion amplitudes.

The results 1-5, on the other hand, elucidate the fact that
we can naturally incorporate massive fermions into the holonomy formalism.
Motivated by these results, in a forthcoming paper, we shall
consider massive deformation of gauge bosons
in the same framework.

\vspace{0.2cm}


\end{document}